\documentclass[fleqn,usenatbib]{mnras}

\usepackage{newtxtext,newtxmath}
\usepackage{float}
\floatstyle{plaintop}
\usepackage[labelfont=bf]{caption}
\restylefloat{table}
\usepackage[T1]{fontenc}
\usepackage{placeins}

\DeclareRobustCommand{\VAN}[3]{#2}
\let\VANthebibliography\thebibliography
\def\thebibliography{\DeclareRobustCommand{\VAN}[3]{##3}\VANthebibliography}

\usepackage{graphicx}	% Including figure files
\usepackage{amsmath}	% Advanced maths commands
\usepackage{hyperref}
\title[The formation of ETGs]{The formation of early-type galaxies through monolithic collapse of gas clouds in Milgromian gravity}
\author[Eappen et al.]{ Robin Eappen$^{1}$\thanks{E-mail: eappen@sirrah.troja.mff.cuni.cz}, Pavel Kroupa$^{1,2}$\thanks{E-mail: pkroupa@uni-bonn.de}, Nils Wittenburg$^{2}$, Moritz Haslbauer$^{2,3}$ and Benoit Famaey$^{4}$
\\
$^{1}$Charles University in Prague, Faculty of Mathematics and Physics, Astronomical Institute, V Holešovickách 2, CZ-180 00 Praha 8, Czech Republic\\
$^{2}$Helmholtz-Institut für Strahlen- und Kernphysik (HISKP), Universität Bonn, Nussallee 14–16, 53115 Bonn, Germany\\
$^{3}$Max-Planck-Institut f\"ur Radioastronomie, Auf dem H\"ugel 69, D-53121 Bonn, Germany\\
$^{4}$Universit´e de Strasbourg, CNRS, Observatoire astronomique de Strasbourg, UMR 7550, F-67000 Strasbourg, France}
\date{Accepted 2022 August 4. Received 2022 August 2; in original form 2021 September 22}
\pubyear{2022}
\begin{document}
\label{firstpage}
\pagerange{\pageref{firstpage}--\pageref{lastpage}}
\maketitle

\begin{abstract}
Studies of stellar populations in early-type galaxies (ETGs) show that the more massive galaxies form earlier and have a shorter star formation history (SFH). In this study, we investigate the initial conditions of ETG formation. The study begins with the collapse of non-rotating post-Big-Bang gas clouds in Milgromian (MOND) gravitation. These produce ETGs with star-forming timescales (SFT) comparable to those observed in the real Universe. Comparing these collapse models with observations, we set constraints on the initial size and density of the post-Big-Bang gas clouds in order to form ETGs. The effective-radius--mass relation of the model galaxies falls short of the observed relation. Possible mechanisms for later radius expansion are discussed. Using hydrodynamic MOND simulations this work thus for the first time shows that the SFTs observed for ETGs may be a natural occurrence in the MOND paradigm. We show that different feedback algorithms change the evolution of the galaxies only to a very minor degree in MOND. The first stars have, however, formed more rapidly in the real Universe than possible just from the here studied gravitational collapse mechanism. Dark-matter-based cosmological structure formation simulations disagree with the observed SFTs at more than 5 sigma confidence.
\end{abstract}

%The aim of this work is to study the initial conditions that forms the early type galaxies in the Universe. The study begins with the collapse of a post-big-bang gas cloud with minimal feedback mechanism under Milgromian (MOND) framework to produce early-type galaxies that are comparable with the star-forming time scales as observed in the real Universe. Comparing these collapse models with the observations we put constraints on the initial conditions for elliptical galaxies in the high redshift universe. This work thus for the first time shows that the star-forming time scales observed for the early-type galaxies is a natural occurrence in the MOND paradigm.


%%%%%%%%%%%%%%%%%%%%%%%%%%%%%%%%%%%%%%%%%%%%%%%%%%

%%%%%%%%%%%%%%%%%%%%%%%%%%%%%%%%%%%%%%%%%%%%%%%%%%
%Notes for Robin
%In the results section, use each figure to explain everything in detail. Use each radii to describe the figure similar to Wit 2020. Then add more literature study to discussions and conclusion as well.
%%%%%%%%%%%%%%%%%%%%%%%%%%%%%%%%%%%%%%%%%%%%%%%%%%

%%%%%%%%%%%%%%%%% BODY OF PAPER %%%%%%%%%%%%%%%%%%

\section{Introduction}
\label{sec:introduction}

Present-day early-type galaxies (ETGs) host old stars without much gas and with a very low star formation activity. Studying the stellar properties of these galaxies is very important for understanding the cosmic history of the Universe. \citet{2017Natur.544...71G} notes that extreme star-formation events in the early Universe are not rare events and that they play a significant role in the early mass assembly of these galaxies. They report the discovery of a quiescent galaxy with stellar mass $\approx 1.7 \times 10^{11} M_{\odot}$ at z = 3.17. \citet{2018MNRAS.475.3700M}, in their stellar population analysis of ETGs, report that the bulk of the stars form in a short timescale with star formation lasting longer in the central regions, and they suggest that an early monolithic formation is highly likely. The observational evidence is reviewed in \citet{2021A&A...655A..19Y} and \citet{2020MNRAS.498.5652K}, who show how the formation of super-massive black holes is a natural outcome of the formation of early-type galaxies within the integrated galaxy-wide IMF (IGIMF, \citealt{2003ApJ...598.1076K,2006MNRAS.365.1333W,2018A&A...620A..39J,2021A&A...655A..19Y}) theory.

%We would also like to note that though the existence of a direct relation between stellar-mass and stellar-ages in a galaxy is evident, there are number of works that shows a dependence of galaxy stellar population ages on galaxy sizes at fixed mass \citep{2009ApJ...698.1232V,2012Natur.484..485C,2013ApJ...762...77P}. 

The observations thus point towards putative progenitors of ETGs undergoing very rapid early formation \citep{1996AJ....112..839C,2005ApJ...621..673T,2005ApJ...632..137N,2009A&A...499..711R,2015MNRAS.448.3484M,2016ApJ...818..179L}. Studies based on elemental abundances of the ETGs indeed constrain them to have formed rapidly on a timescale of $\approx$ 1 Gyr. Also, studies from stellar population models with observed line indices show that ETGs must have formed the bulk of their stars in a short timescale. \citet{1996AJ....112..839C}, suggested the name "downsizing" to describe that the less massive galaxies have more extended SFHs compared to the massive ones.

The duration of the formation, i.e. the star-forming timescale (SFT) for ETGs is expressed in \citet{2005ApJ...621..673T}, \citet{2009A&A...499..711R} and \citet{2015MNRAS.448.3484M} in terms of the downsizing time which is a function of the mass of the galaxy. These downsizing timescales correspond to approximately 0.34 $\mathrm{Gyr}$ for a galaxy with a present-day baryonic mass of $10^{12}M_{\rm \odot}$, suggesting that it should have formed under a monolithic cloud collapse scenario rather than hierarchical merging which requires longer time-spans \citep{2010ApJ...722.1666W,2010MNRAS.406..230R}. On average, more massive galaxies have been found to have formed the bulk of their stars on a shorter timescale and have completed their star-formation activity at a higher redshift \citep{2005ApJ...626..680D,2008ApJ...676..781W,2012A&A...537A..31C} which points to a strong dependence of the SFT in a galaxy on its present-day stellar-mass content.

Massive ETGs are thought to be important tracers of the cosmic history of the stellar mass assembly and of galaxy evolution. A theoretical model that describes ETGs should explain how and why they have these observed SFTs. The standard model of cosmology (SMoC) has been successful in forming ETGs  (see \citealt{2017ARA&A..55...59N} for a detailed review on the theory and numerical simulations in the SMoC) but has not been able to reproduce ETGs with SFTs (see Fig. \ref{fig:tngtau}) similarly short as the observed ETGs \citep{2005ApJ...621..673T,2009A&A...499..711R,2015MNRAS.448.3484M}. With the existence of dark matter (DM) remaining a hypothesis, an alternate theory of cosmology should be considered \citep{2012IJMPD..2130003K,2015CaJPh..93..169K}.

%A future work would further investigate the morphology and structure of these collapse models with more complex feedback mechanisms and factor in the galaxy-wide IMF that would account for the expansion as observed in the real universe.%
By combining observational constraints on the dynamics in galaxies with constraints from the Solar System, \cite{1983ApJ...270..365M} corrected the theory of gravitation at low acceleration, which could be a consequence of the quantum vacuum \citep{1999PhLA..253..273M,2013PhRvD..87h4063P,2017ScPP....2...16V,PhysRevD.96.083523}.  In analogy with Newtonian dynamics, a non-relativistic Milgromian gravity theory (MOND) can be constructed by setting up a Lagrangian which, upon extremization of the action, yields a generalised Milgromian Poisson equation. Two such Lagrangians have been proposed, called AQUAL \citep{1984ApJ...286....7B} and QUMOND \citep{2010MNRAS.403..886M}. The latter, which we adopt hereafter, yields the following generalized Poisson equation,
\begin{equation} 
%\nabla \cdot [\mu(\left |\nabla \Phi \right |/a_0)\nabla \Phi] = 4 \pi G \rho,
\Delta \Phi(\Vec{x}) = 4\pi G\rho_{\rm b} (\Vec{x}) + \Vec{\nabla} \cdot [\tilde{v}(\left\lvert\vec{\nabla} \phi \right\rvert/a_0)\Vec{\nabla} \phi (\Vec{x})],
\label{eq:mil1}
\end{equation}
or,
\begin{equation} 
\Delta \Phi(\Vec{x}) = 4\pi G [\rho_{\rm b} (\Vec{x}) + \rho_{\rm ph}(\Vec{x})],
\label{mil2}
\end{equation}
where $\rm \rho_{\rm b}(\it\vv{x})$ is the baryonic density, $\rm \phi(\it\vv{x})$ is the Newtonian potential which fulfils the standard Poisson equation, $\rm \Delta \phi(\it\vv{x})$ = 4$\rm \pi$G$\rm \rho_{\rm b}(\it\vv{x})$ and Milgrom's constant $a_{\rm 0}$ $\approx \rm 1.2 \times 10^{-10} \ m\ s^{-2} \approx 3.8 \ pc\ Myr^{-2}$ . The phantom dark matter (PDM) density, $\rm \rho_{ph}(\it\vv{x})$, is not a real density distribution but a mathematical function that arises out of the non-linearity of the Poisson equation. $\Phi(\it\vv{x})$ is the total gravitational potential from which the accelerations follow, $\vv{a} = -\vv{\nabla} \vv{\Phi}$, and $\tilde{v}(y)$ is a transition function characterizing the theory (see \citealt{2008arXiv0801.3133M,2010MNRAS.403..886M,2014SchpJ...931410M,2012LRR....15...10F} and \citealt{2022Symm...14.1331B} for detailed reviews on the theory). $\tilde{v}(y)$ has the limits,
\begin{equation} 
\tilde{v}(y) \rightarrow 0 \;\rm{for}\; \it y\; \gg \rm 1 \;\rm{and}\; \it \tilde{v}(\it y)\; \rightarrow \it y^{\rm -1/2}\; \rm{for}\; \it y\; \ll \rm 1.
\label{mil01}
\end{equation}
The above formulation deals only with linear differential equations and is shown to emerge as a natural modification of a Palatini-type formulation of Newtonian gravity. It is a member of a larger class of bi-potential theories (quasi-linear formulation of MOND, QUMOND \citealt{2010MNRAS.403..886M}). \citet{1984ApJ...286....7B} also noted a correspondence with some theories of quark confinement using a different form of the function $\tilde{v}$.

MOND predicted galaxy scaling relations obeyed by galaxies \citep{1983ApJ...270..371M} such as the Baryonic Tully Fisher Relation (BTFR) \citep{2000ApJ...533L..99M,2005ApJ...632..859M,2012AJ....143...40M,2016ApJ...818..179L} and the Radial Acceleration Relation (RAR) \citep{1990A&ARv...2....1S,2004ApJ...609..652M,2017ApJ...836..152L}.

Any realistic theory of galaxy formation has to address the question of when and how the stars form in them. There exists a large body of literature on the formation of ETGs in the SMoC \citep{2017ARA&A..55...59N} for a review. Here we concentrate on the question whether ETGs can form in Milgromian gravitation from collapsing post-Big-Bang gas clouds and how these models compare to the observed ETGs. The aim of this work is to use simulations of collapsing and non-rotating post-Big-Bang gas clouds to study the duration of their SFTs and the resulting size--mass relation. Galaxy scale fluctuations in a MOND Universe should grow by the application of MOND to the peculiar accelerations allowing for large density contrasts and for large galaxies to form as early as $z = 10-25$ from the collapse of almost isolated gas clouds. \citet{2007ApJ...660..256N} and \citet{2008MNRAS.386.1588S} have therefore studied disipationless collapse in a MOND context. Here we model for the first time the collapse of massive non-rotating gas clouds including the physics of star-formation. The results are found to be very close to the observations. Previous work has shown the important bulk properties of disk galaxies to follow naturally from the collapse of initially rotating post-Big-Bang gas clouds \citep{2020ApJ...890..173W}.

The layout of this paper is as follows. In Section \ref{sec:nummethod}, the numerical hydro-dynamical code used is discussed briefly and in Section \ref{sec:models}, the details of the simulations are described. The results from the work are presented in Section \ref{sec:results}. Section \ref{sec:discussion} contains the discussion and Section \ref{sec:Conclusion} has conclusions along with potential future work.
%%%%%%%%%%%%%%%%%%%%%%%%%%%%%%%%%%%%%%%%%%%%%%%%%%
\section{Phantom Of RAMSES (POR)}
\label{sec:nummethod}
The Phantom of RAMSES (POR) code \citep{2015CaJPh..93..232L,2021CaJPh..99..607N} is a customized version of RAMSES \citep{2002A&A...385..337T} applying the adaptive mesh refinement (AMR) method to solve the Milgromian Poisson equation (Eq. \ref{eq:mil1}). It uses a multi-grid and a conjugate gradient solver to solve the generalised Poisson equation \footnote{\citet{2015MNRAS.446.1060C} developed a similar code independently called RAyMOND.}.

There are several transition functions used in the literature to accommodate the non-linearity of the Poisson equation in MOND. POR uses, 
\begin{equation} 
\tilde{v} (y) = -\frac{1}{2} + \left( \frac{1}{4}+\frac{1}{y}\right)^\frac{1}{2},
\label{eq:mil3}
\end{equation}
where $y =|\vv{\nabla} \phi|/a_{0}$ (see \citealt{2015CaJPh..93..232L}).

The POR working scheme involves first solving the standard Poisson equation to compute the Newtonian potential, and then the PDM density is calculated using a discrete scheme (see \citealt{2015CaJPh..93..232L}). Then, the Poisson equation is solved for a second time with the Newtonian density and PDM density to compute the total gravitational potential.

As for now, POR has been successfully applied in the simulations of Antennae-like galaxies \citep{2016MNRAS.463.3637R}, simulations of the Sagittarius satellite galaxy \citep{2017A&A...603A..65T}, simulations of the Local Group producing the planes of satellites \citep{2018A&A...614A..59B,2021Galax...9..100B,2022MNRAS.513..129B}, simulations of streams from globular clusters \citep{2018A&A...609A..44T}, simulations of the formation of exponential disk galaxies \citep{2020ApJ...890..173W}, the global stability of M33 \citep{2020ApJ...905..135B}, the evolution of globular-cluster systems of ultra-diffuse galaxies due to dynamical friction \citep{2021arXiv210705667B} and polar-ring galaxies have also been successfully modelled with a pre-POR code \citep{2013MNRAS.432.2846L}. A detailed user-guide can be found in \citet{2021CaJPh..99..607N}.
%%%%%%%%%%%%%%%%%%%%%%%%%%%%%%%%%%%%%%%%%%%%%%%%%%
\section{Models}
\label{sec:models}
We use the POR numerical code introduced in Section \ref{sec:nummethod} to set-up post-Big-Bang gas clouds in an isolated environment in-order to allow a first assessment of how ETGs would have formed in MOND. Here, we start with initial conditions comparable to those in \citet{2020ApJ...890..173W}, but with no initial rotation.

Stellar particles are created in POR if the gas mass–volume–density exceeds a user-defined threshold value. The code checks at every time step whether any cell exceeds the density threshold, and then a stellar particle is created. Details on the conditions for the star formation threshold value and various other code parameters are described in \citet{2015CaJPh..93..232L}, \citet{2020ApJ...890..173W} and \citet{2021CaJPh..99..607N}.

\begin{table*}
\centering
	\caption{The initial conditions and results of all the models.}
	\label{tab:init_cond}
%\centering
\begin{tabular}{c c c c c c c c}
   
		\hline
		Model & $M_{\rm initial}$ & $R_{\rm initial}$ & $\Delta\tau_{\rm m}$ & $t_{\rm start}$ & $t_{\rm peak}$ & $r_{\rm eff}$ & $\it{SFR}_{\rm peak}$\\
		Name & (10$^9$ $M_\odot$) & (kpc) & (Gyr) & (Gyr) & (Gyr) & (kpc) & ($M_{\odot}\rm yr^{-1}$)\\
        \hline
e1 & 0.6 & 50 & 0.62 & 1.18 & 2.15 & 0.43 & 6.71E+00 \\
e2 & 1 & 50 & 0.53 & 1.02 & 1.86 & 0.47 & 1.34E+01 \\
e3 & 6.4 & 50 & 0.17 & 0.51 & 0.82 & 0.39 & 2.66E+02 \\
e4 & 10 & 50 & 0.13 & 0.44 & 0.71 & 0.6 & 5.20E+02 \\
e5 & 30 & 50 & 0.07 & 0.29 & 0.49 & 0.84 & 2.06E+03 \\
e6 & 50 & 50 & 0.05 & 0.23 & 0.44 & 0.73 & 4.40E+03 \\
e7 & 64 & 50 & 0.05 & 0.21 & 0.41 & 0.67 & 7.22E+03 \\
e8 & 70 & 50 & 0.04 & 0.2 & 0.39 & 0.77 & 7.94E+03 \\
e9 & 100 & 50 & 0.04 & 0.17 & 0.35 & 0.83 & 1.45E+04 \\
e10 & 0.6 & 100 & 0.91 & 2.46 & 3.5 & 0.4 & 4.88E+00 \\
e11 & 1 & 100 & 0.78 & 2.11 & 3.95 & 0.53 & 1.03E+01 \\
e12 & 5 & 100 & 0.37 & 1.19 & 1.73 & 0.47 & 9.67E+01 \\
e13 & 10 & 100 & 0.26 & 0.95 & 1.41 & 0.47 & 2.77E+02 \\
e14 & 20 & 100 & 0.19 & 0.73 & 1.21 & 1.41 & 6.05E+02 \\
e15 & 30 & 100 & 0.16 & 0.63 & 1.07 & 0.56 & 1.12E+03 \\
e16 & 50 & 100 & 0.13 & 0.52 & 0.92 & 0.68 & 2.52E+03 \\
e17 & 64 & 100 & 0.11 & 0.48 & 0.87 & 0.75 & 3.97E+03 \\
e18 & 70 & 100 & 0.11 & 0.46 & 0.85 & 0.77 & 4.40E+03 \\
e19 & 100 & 100 & 0.09 & 0.4 & 0.76 & 0.99 & 7.78E+03 \\
e20 & 0.6 & 200 & 1.62 & 5.12 & 6.4 & 0.4 & 3.15E+00 \\
e21 & 1 & 200 & 1.38 & 4.37 & 7.57 & 0.56 & 6.38E+00 \\
e22 & 10 & 200 & 0.53 & 1.95 & 2.98 & 0.55 & 1.27E+02 \\
e23 & 30 & 200 & 0.34 & 1.31 & 2.18 & 0.68 & 6.25E+02 \\
e24 & 50 & 200 & 0.28 & 1.11 & 1.90 & 1.21 & 1.32E+03 \\
e25 & 70 & 200 & 0.25 & 0.99 & 1.76 & 0.93 & 2.18E+03 \\
e26 & 100 & 200 & 0.22 & 0.87 & 1.59 & 0.93 & 3.62E+03 \\
e27 & 0.6 & 300 & 1.65 & 7.91 & 9.29 & 0.31 & 2.54E+00 \\
e28 & 5 & 300 & 0.79 & 3.81 & 5.09 & 0.46 & 4.51E+01 \\
e29 & 10 & 300 & 0.65 & 2.99 & 4.48 & 1.03 & 9.41E+01 \\
e30 & 30 & 300 & 0.48 & 2.02 & 3.31 & 0.76 & 4.48E+02 \\
e31 & 50 & 300 & 0.42 & 1.68 & 2.9 & 0.73 & 8.96E+02 \\
e32 & 70 & 300 & 0.39 & 1.49 & 2.66 & 0.83 & 1.38E+03 \\
e33 & 100 & 300 & 0.35 & 1.32 & 2.42 & 1.36 & 2.34E+03 \\
e34 & 6.4 & 500 & 1.41 & 5.97 & 8.44 & 0.41 & 3.29E+01 \\
e34c & 6.4 & 500 & 1.43 & 5.96 & 8.41 & 0.39 & 3.25E+01 \\
e35 & 10 & 500 & 1.18 & 5.09 & 7.45 & 0.47 & 6.41E+01 \\
e36 & 30 & 500 & 0.75 & 3.43 & 5.58 & 0.6 & 2.85E+02 \\
e37 & 50 & 500 & 0.62 & 2.85 & 4.88 & 0.84 & 5.95E+02 \\
e38 & 64 & 500 & 0.55 & 2.58 & 4.48 & 0.95 & 8.60E+02 \\
e39 & 70 & 500 & 0.54 & 2.48 & 4.4 & 1.04 & 9.39E+02 \\
e39c & 70 & 500 & 0.53 & 2.47 & 4.44 & 1.09 & 9.48E+02 \\
e40 & 1000 & 500 & 0.24 & 1.03 & 1.9 & 1.7 & 1.48E+04 \\
        \hline
    %\multicolumn{3}{l}{where M$\rm _i$ is the initial mass of the cloud, R$\rm _i$ is the initial radius of the cloud, $\rm \tau_m$ is the star-forming timescale for the models, age$\rm _m$ is the age of the models and, r$\rm _e$ is the half mass radius of the models.}
%Model Name, is the name of each models, M$_i$ is the initial mass of the cloud and R$_i$ is the initial radius of the cloud\\
    \end{tabular}
\\Note: $M_{\rm initial}$ is the initial mass of the post-Big-Bang gas cloud, $R_{\rm initial}$ is the initial radius of the post-Big-Bang gas cloud, $\Delta\tau_{\rm m}$ is the resulting SFT, $t_{\rm peak}$ is the time after start of the simulation when the first stellar particle is formed, $t_{\rm peak}$ is the resulting time when the peak of the SFH is observed in the simulation, $r_{\rm eff}$ is the resulting projected effective-radius and $\it{SFR}_{\rm peak}$ is the resulting SFR at the peak of the SFH. All the models here have a maximum resolution of 0.24 kpc except for models e34c and e39c which have a maximum resolution of 0.06 kpc and 0.12 kpc, respectively. Models e34c and e39c are computations with complex feedback (see Section \ref{sec:complexfeedback}).
\end{table*}

In total 42 different model galaxies were calculated for this study to understand how different initial masses, $M_{\rm initial}$, and initial radii, $R_{\rm initial}$, of the gas cloud can affect the SFT (Table \ref{tab:init_cond}). Only the simple cooling/heating feedback algorithm is used in the simulation of all the model galaxies except models e34c and e39c, where a more complex feedback algorithm is used (see Section \ref{sec:complexfeedback}). The computations are run for 10 Gyr. The starting temperature of the gas is $\mathrm{10^{4} K}$ and the size of the simulation box is 1000 kpc. The minimum and maximum refinement levels for the models \footnote{Models e34c and e39c have maximum refinement levels of 14 and 13, respectively, which sets the maximum spatial resolution to 0.06 kpc and 0.12 kpc.} are 7 and 12 respectively which sets a limit to the minimum spatial resolution of 7.81 kpc and maximum spatial resolution of 0.24 kpc, respectively, for each simulation \citep{2002A&A...385..337T,2015CaJPh..93..232L}. 
\begin{figure}
	\includegraphics[width=\columnwidth]{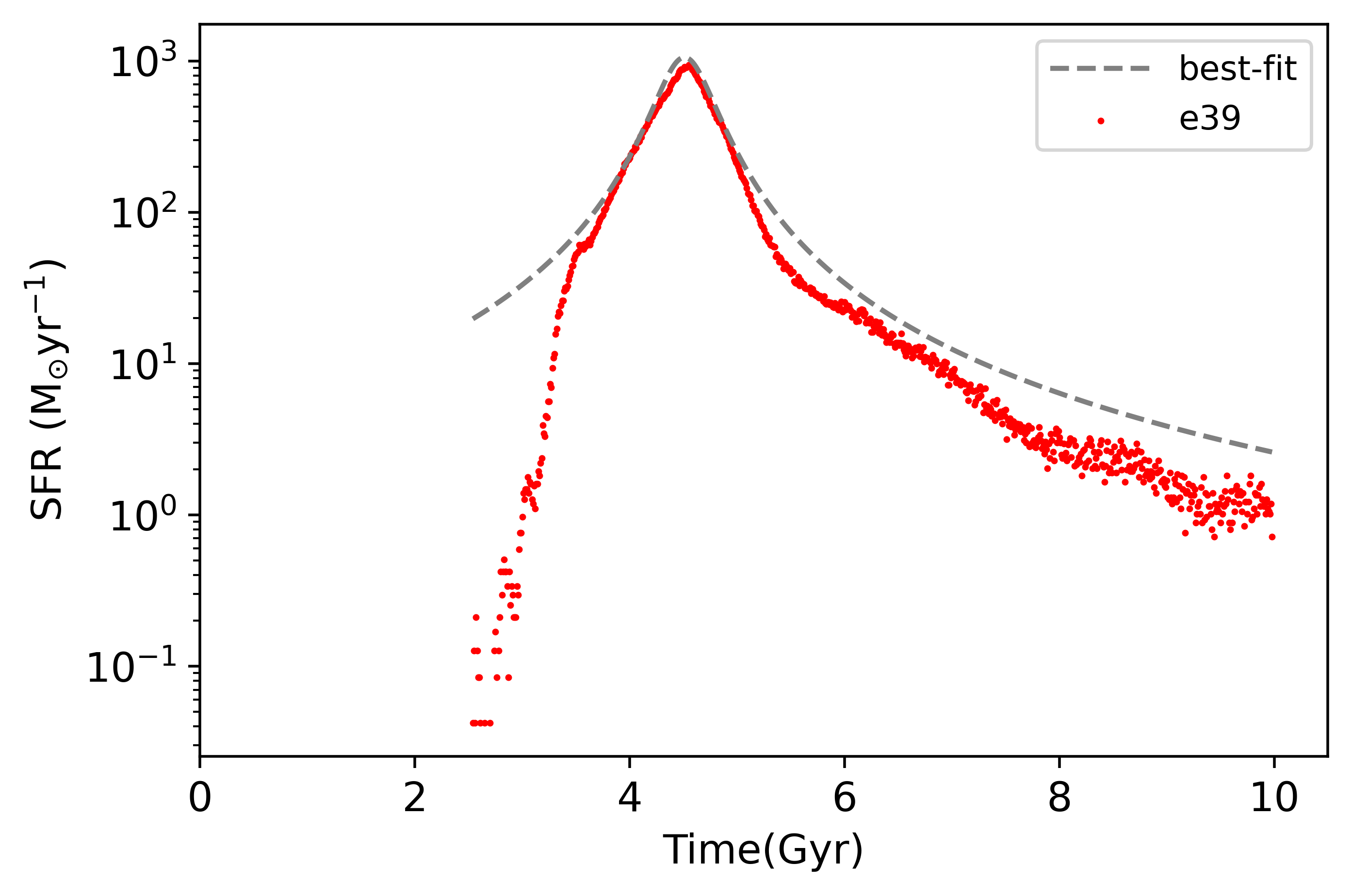}
     \caption{SFH of model e39 (Table \ref{tab:init_cond}). The red dots are the SFR at each $\delta t$ and the dashed grey line is the calculated best-fit (Eq. \ref{eq:sfhfit}). The x-axis is the time in the computational domain where the computation and thus the collapse starts at 0 and the calculation ends at 10 Gyr. The fit is done only to obtain the FWHM of the SFH and the discrepancy at large $t$ is not important for the purpose of this study.}
    \label{fig:sfhplot}
\end{figure}

In all computations, the star formation rate (SFR) is calculated by separating all stellar particles in bins of $\delta t$ = 10 Myr according to their age, summing up the stellar mass in every time bin and dividing this by the length of $\delta t$. The SFR increases sharply at the time of the collapse of the sphere until the maximum is reached and then it decreases approximately exponentially (Fig. \ref{fig:sfhplot}). The SFH of a galaxy is constructed by the consecutive SFRs of each $\delta t$ time step. The SFH can be represented by a Lorentz function (dashed grey line in Fig. \ref{fig:sfhplot}),
%\begin{equation}
\begin{multline}
    \it{SFR}/{M_{\odot}\mathrm{yr^{-1}}} = \frac{\it{SFR}_{\mathrm{peak}}/M_{\odot}\mathrm {yr^{-1}}}{\mathrm{\pi}} + \\ \left( \frac{0.5\cdot\Delta\tau_{\mathrm{m}}/\mathrm{Gyr}}{(t/\mathrm{Gyr}-t_{\mathrm{peak}}/\mathrm{Gyr})^{2}+(0.5\cdot\Delta\tau_{\mathrm{m}}/\mathrm{Gyr})^{2}}\right),
    \label{eq:sfhfit}
\end{multline}
%\end{equation}
where $t_{\rm peak}$ is the time in the simulation where the peak of the SFH is observed and $\Delta\tau_{\rm m}$ is the SFT for the model. The full-width at half maximum (FWHM) of the SFH gives the duration for which the bulk of the star formation takes place, so we use the FWHM to define the SFT (i.e. $\Delta\tau_{\rm m}$) for the models in this work.

The collapse for models with the same $M_{\rm initial}$ and varying $R_{\rm initial}$ shows that $\Delta\tau_{\rm m}$ increases with increasing $R_{\rm initial}$. The sum of all formed stellar particles at the end of each simulation is the final stellar mass \footnote{We do not list the $M_{\rm final}$ since the final mass of the model galaxy is equal to the initial mass of the gas cloud.}, $M_{\rm final}$, of each model, independently of the chosen feedback algorithm. The models here do not loose mass through outflows.

We also calculate the time taken for the first stellar particle to form, $t_{\rm start}$, and compare it to the MOND free-fall time,
\begin{equation}
    t_{\mathrm{ff,mond}} = \left(\frac{\pi}{2}\right)^{0.5} \cdot \frac{R_{\mathrm{initial}}}{(G \cdot M_{\mathrm{initial}} \cdot a_{\rm 0})^{0.25}} ,
	\label{eq:mondfreefall}
\end{equation}
(eq. 27 in \citealt{2021MNRAS.506.5468Z}) in Fig. \ref{fig:mondfftime}. The free-fall time can be used as an approximation for the time the gas needs to collapse into stars. The models have $t_{\rm start}$ comparable to the MOND free-fall time (Fig. \ref{fig:mondfftime}).
%%%%%%%%%%%%%%%%%%%%%%%%%%%%%%%%%%%%%%%%%%%%%%%%%%
\section{Results}
\label{sec:results}
We compare the SFTs, ages, and effective-radii of our simulated galaxies with the observed galaxy properties in this section. The average ages estimated by \citet{2005ApJ...621..673T} from absorption line indices of 124 ETGs, $t_{\rm thomas}$, for both high-density and low-density environments (quantities in brackets are the values derived for the high-density environment) can be written,
\begin{equation}
    t_{\mathrm{thomas}} = 1.33(2.67) \cdot (M_{\mathrm{final}}/M_{\mathrm{\odot}})^{0.07(0.05)},
	\label{eq:thomasage}
\end{equation}
(eq. 3 in \citealt{2005ApJ...621..673T}) and the relation deduced for the SFTs is,
\begin{equation}
    \Delta\tau_{\rm thomas}/\mathrm{Gyr} = 3.67 - 0.37 \cdot \mathrm{log_{10}}(M_{\mathrm{final}}/M_{\mathrm{\odot}})
	\label{eq:thomasSFT}
\end{equation}
(eq. 5 in \citealt{2005ApJ...621..673T}), being the same for high- and low-density environments.

%%%%%%%%%%%%%%%%%%%%%%%%%%%%%%%%%%%%%%
\subsection{Downsizing SFH}
\label{sec:SFT1}
\begin{figure}
	\includegraphics[width=\columnwidth]{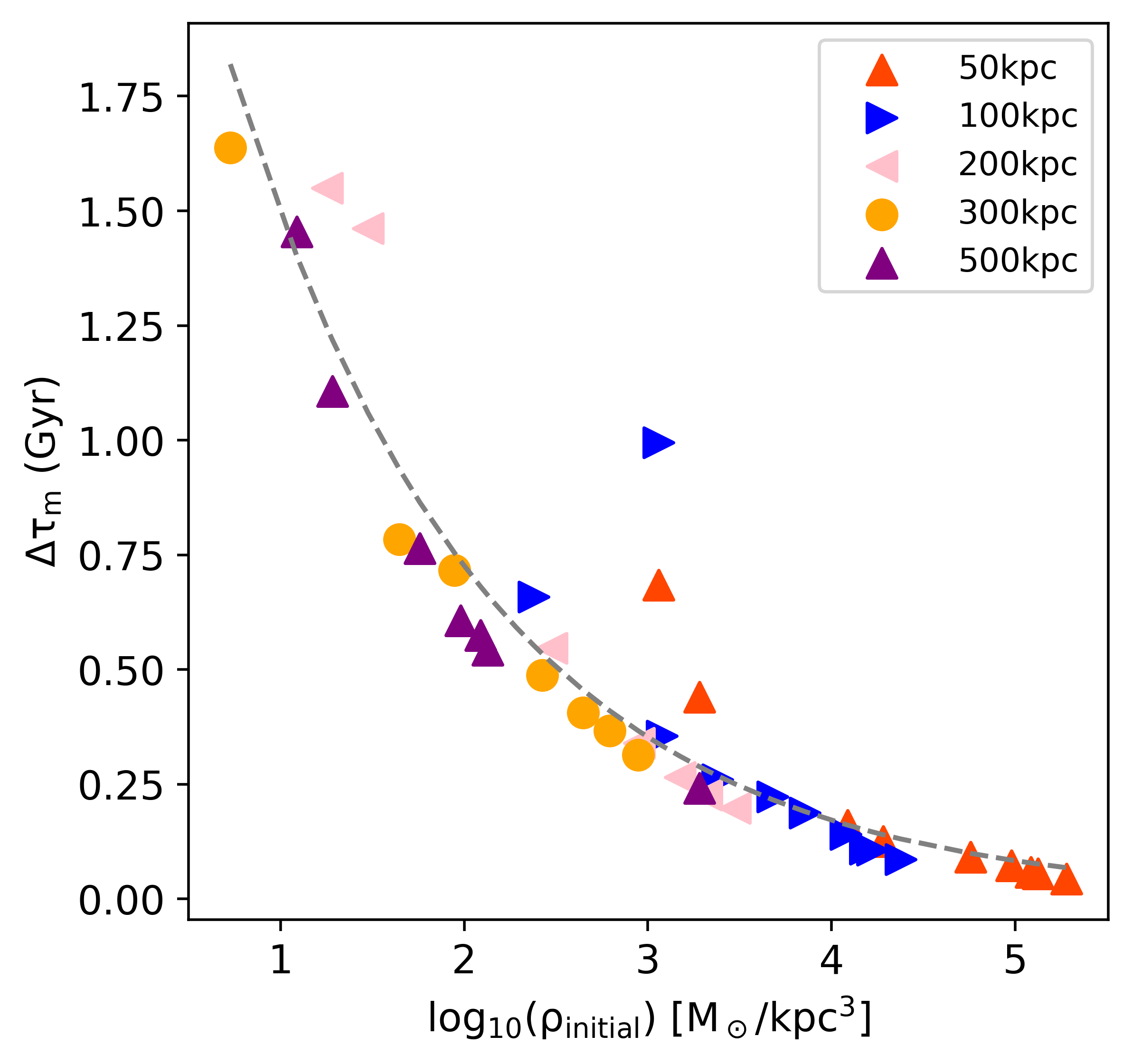}
    \caption{SFT--density relation for all models listed in Table \ref{tab:init_cond}. The coloured markers are model galaxies and the dashed grey line is the calculated best-fit (Eq. \ref{eq:dentauscaling}).}
    \label{fig:denvtaumodels}
\end{figure}

The SFTs are found to follow a similar downsizing behaviour as documented in \citet{2005ApJ...621..673T,2010MNRAS.404.1775T}, where, for a given $R_{\rm initial}$, the galaxies with larger masses (higher density) form at an earlier epoch compared to galaxies with lower masses (lower density). Fig. \ref{fig:denvtaumodels} shows that the model galaxies with high initial cloud density, $\rho_{\rm initial}$, form earlier and quicker (shorter $\Delta\tau_{\rm m}$) than model galaxies with low $\rho_{\rm initial}$. We use a power law function to obtain the scaling relation in Fig. \ref{fig:denvtaumodels},
\begin{equation}
    \Delta\tau_{\rm m}/\mathrm{Gyr} = 3.03 \cdot (\rho_{\rm initial}/M_{\odot}\mathrm{kpc^{-3}})^{-0.30}.
	\label{eq:dentauscaling}
\end{equation}

%\textcolor{red}{We also note the time taken for the first stellar particle to form, $t_{\rm start}$ and compare it to the MOND free-fall time (\citealt{2021MNRAS.506.5468Z}, their Eq. 27) in Fig. \ref{fig:timevden}. The models have $t_{\rm start}$ comparable to MOND free-fall time.}
\begin{figure}
	\includegraphics[width=\columnwidth]{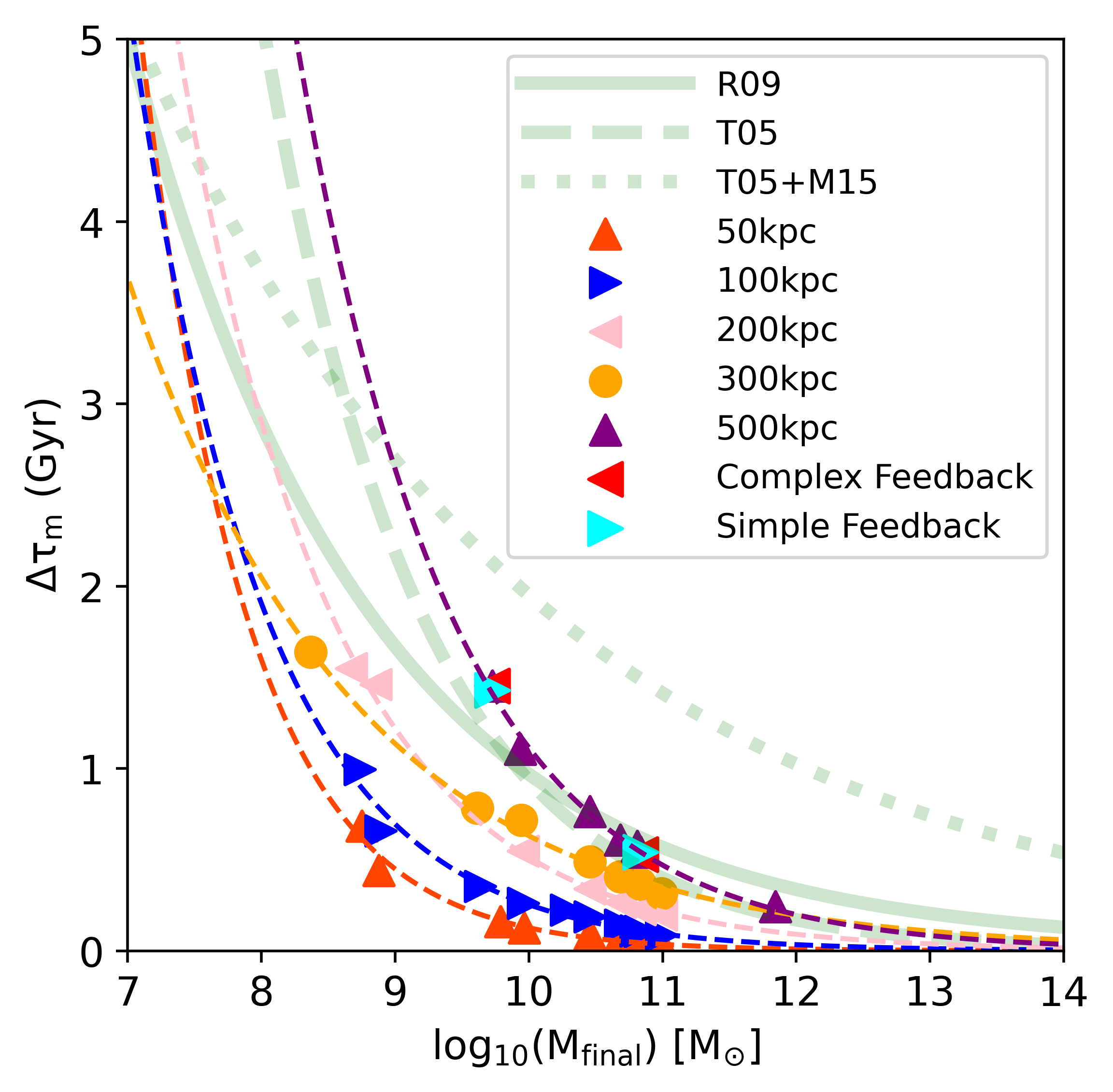}
    \caption{SFT--mass relation. The coloured markers are model galaxies listed in Table \ref{tab:init_cond} with different $R_{\rm initial}$ as indicated by the legend. The observationally constrained relations are from \citet{2005ApJ...621..673T} (dashed green line), \citet{2009A&A...499..711R} (solid green line) and \citet{2021A&A...655A..19Y} (which is a combination of eq. 3 of \citealt{2005ApJ...621..673T} and eq. 3 of \citealt{2015MNRAS.448.3484M}, dotted green line). The red left-pointing triangles are the models with complex feedback (see Section \ref{sec:complexfeedback}).}
    \label{fig:downsizing}
\end{figure}

A downsizing SFH can be well explained using Fig. \ref{fig:downsizing} where the SFT is plotted against $M_{\rm final}$ for each model galaxy along with the observations from \citet{2005ApJ...621..673T}, \citet{2009A&A...499..711R} and \citet{2015MNRAS.448.3484M}.

%\begin{figure}
%	\includegraphics[width=\columnwidth]{figures/bestfitsmodels.png}
%    \caption{Similar to Fig \ref{fig:downsizing} but with just the models in the work and the dotted line is the best-fit for each model.}
%    \label{fig:bestfitsdownsizing}
%\end{figure}
We use a power law function to fit the models and obtain the scaling relation for the models in Fig. \ref{fig:downsizing},
\begin{equation}
    \Delta\tau_{\rm m}/\mathrm{Gyr} = A \cdot (M_{\rm final}/M_{\odot})^{B},
	\label{eq:50kpc}
\end{equation}
where $A$ and $B$ are  fit parameters for each $R_{\rm initial}$ given in Table \ref{tab:downtable1}. From Fig. \ref{fig:downsizing}, it can be seen that models with $R_{\rm initial}$ = 500 kpc lead to similar $\Delta \tau_{\rm m}$ values as the observational results from \citet{2005ApJ...621..673T} and models with $R_{\rm initial}$ = 200 kpc and $R_{\rm initial}$ = 300 kpc are similar to the observational data from \citet{2009A&A...499..711R} but steeper. The models with $R_{\rm initial}$ = 50 kpc and $R_{\rm initial}$ = 100 kpc do not agree with any of the observed relations. We note that our models do not agree with the new downsizing relation obtained by \citet{2021A&A...655A..19Y} and this is discussed in Section \ref{sec:discussion}.
%%%
%%%\begin{equation}
%%%    \rm log_{10} (\tau_{m}/Gyr)_{100kpc} = 3.7855 - 0.4382 \times log_{10} (M_{f}/M_{\odot}) 
%%%	\label{eq:100kpc}
%%%\end{equation}
%%%\begin{equation}
%%%    \rm log_{10} (\tau_{m}/Gyr)_{200kpc} = 3.4756 - 0.3766 \times log_{10} (M_{f}/M_{\odot})
%%%	\label{eq:200kpc}
%%%\end{equation}
%%%\begin{equation}
%%%    \rm log_{10} (\tau_{m}/Gyr)_{300kpc} = 2.3577 - 0.2558 \times log_{10} (M_{f}/M_{\odot}) 
%%%	\label{eq:300kpc}
%%%\end{equation}
%%%\begin{equation}
%%%    \rm log_{10} (\tau_{m}/Gyr)_{500kpc} = 3.7961 - 0.3749 \times log_{10} (M_{f}/M_{\odot})
%%%	\label{eq:500kpc}
%%%\end{equation}
%%%
%%%%%%%%%%%%%%%%%%%%%
\begin{figure}
	\includegraphics[width=\columnwidth]{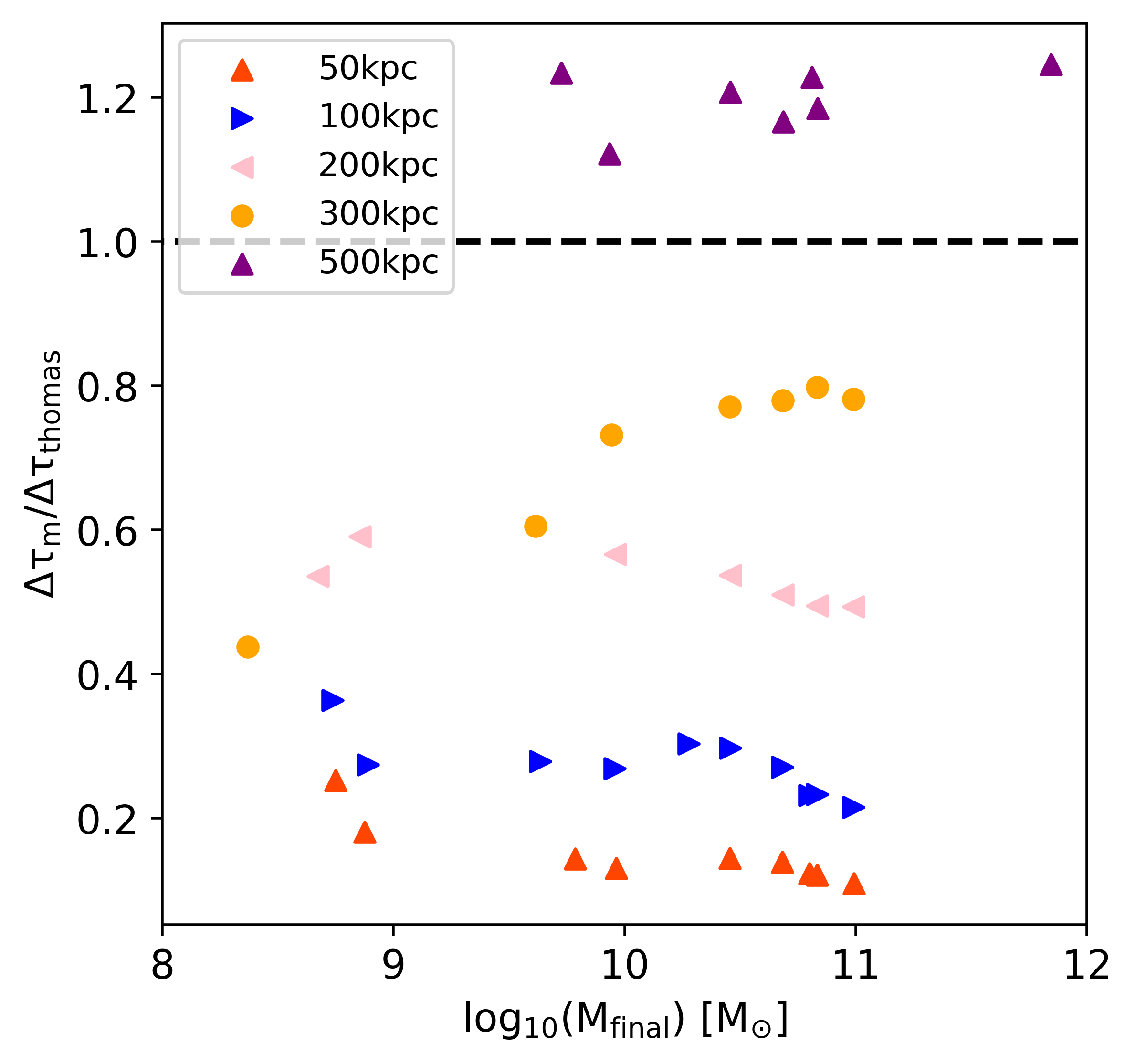}
    \caption{A plot showing the ratio of SFTs of the model galaxies and observed galaxies ($\Delta\tau_{\rm m}/\Delta\it\tau_{\rm{thomas}}$) vs final stellar mass of the galaxy. The coloured markers are model galaxies as listed in Table \ref{tab:init_cond} with different $R_{\rm initial}$ as indicated by the legend. The dashed black line is where $\Delta\tau_{\rm m}$ = $\Delta\tau_{\rm{thomas}}$.}
    \label{fig:bestdown}
\end{figure}

\begin{figure*}
	\includegraphics[width=\columnwidth]{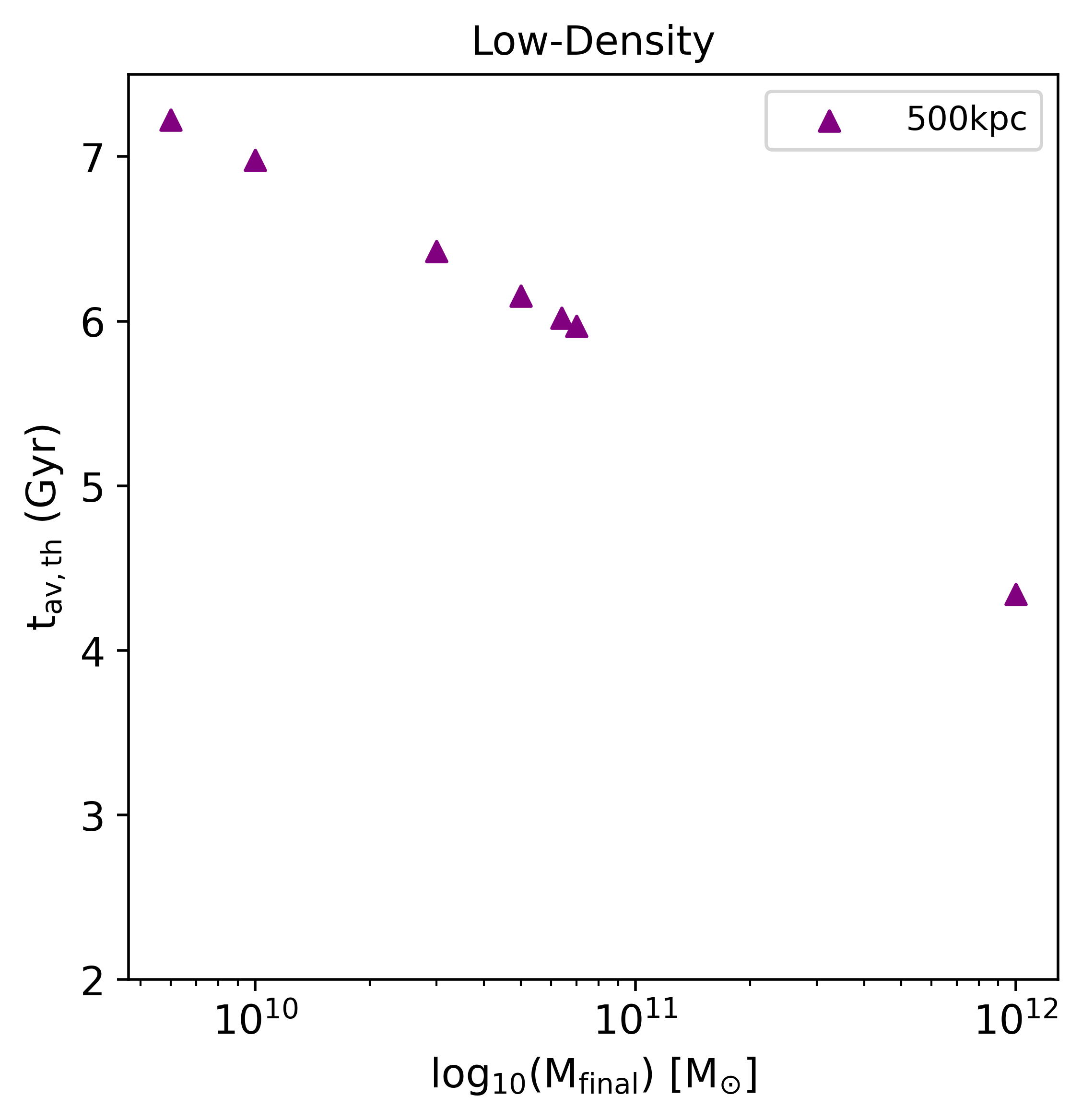}
	\includegraphics[width=\columnwidth]{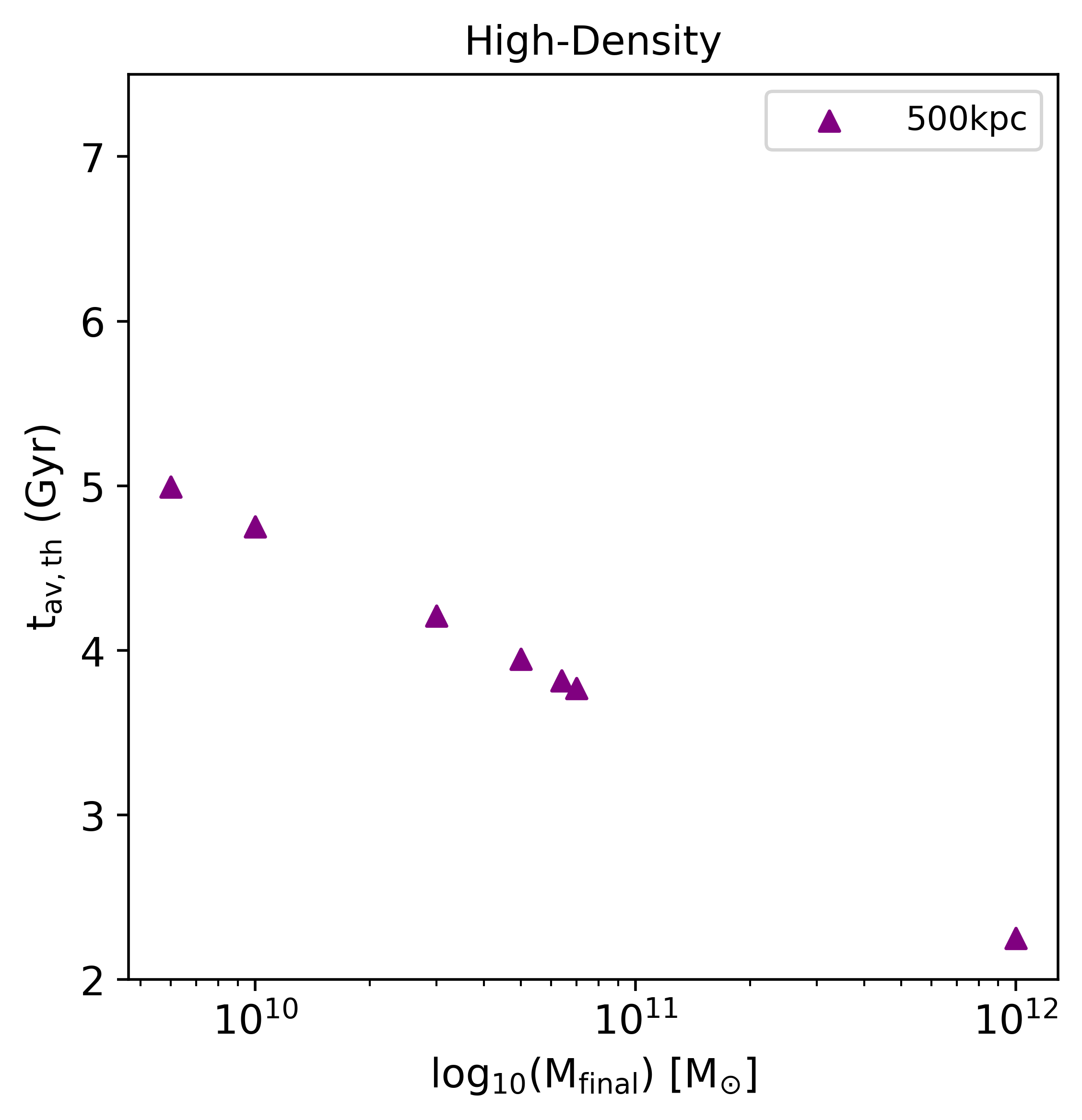}
	\\
	\includegraphics[width=\columnwidth]{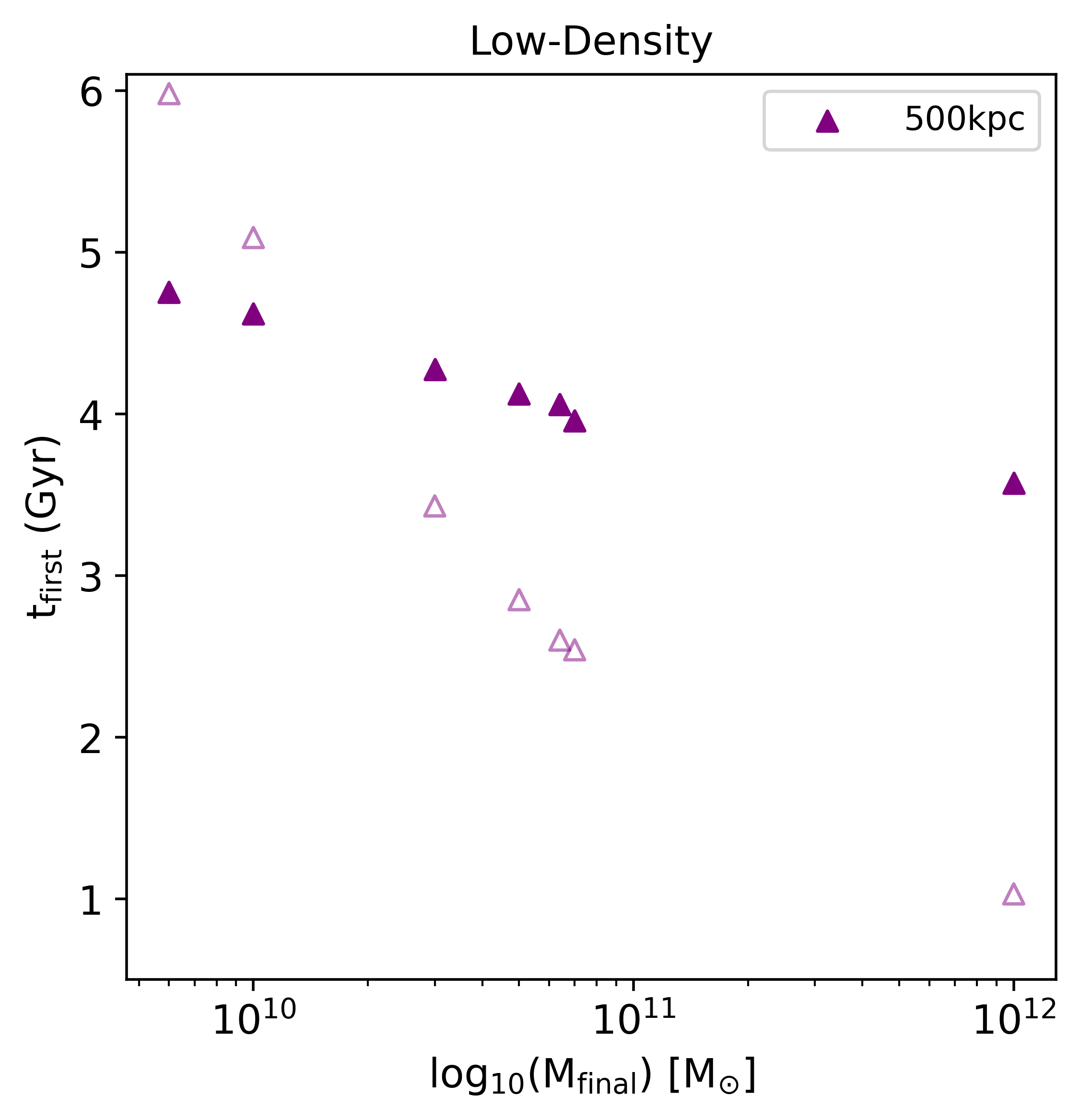}
	\includegraphics[width=\columnwidth]{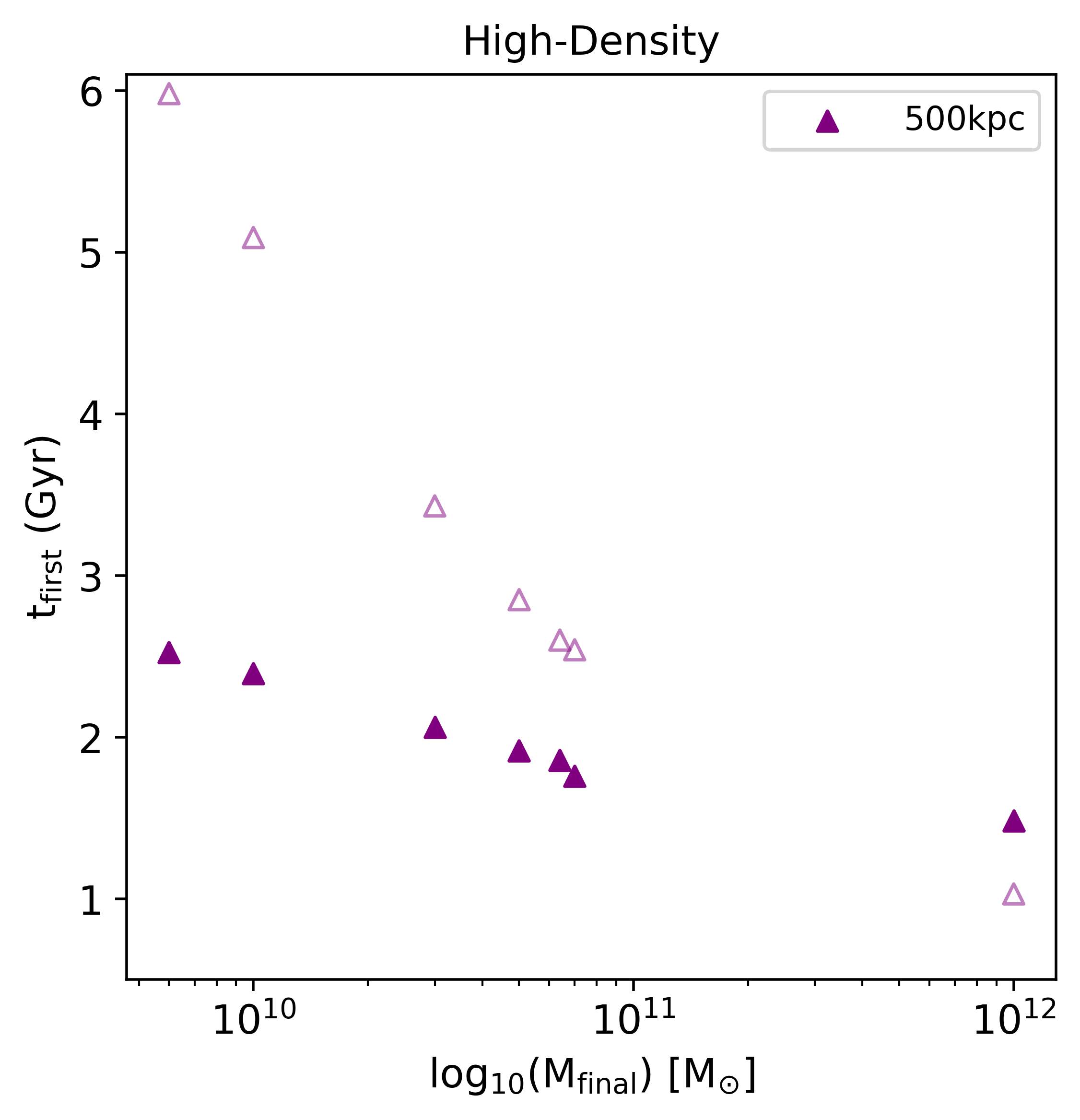}
    \caption{$t_{\rm av,th}$ and $t_{\rm first}$ vs. $M_{\rm final}$ for models with $R_{\rm initial}$ = 500 kpc in the low-density (\emph{left}) and high-density (\emph{right}) environment (Eq. \ref{eq:tshift} $\&$ \ref{eq:tfirst}). The times are computed since the nominal Big-Bang. The unfilled upward pointing purple triangles in the second panel are relative to the start of the computation at $t$ = 0.}
    \label{fig:tavvmfinal}
\end{figure*}

\begin{figure*}
	\includegraphics[width=\columnwidth]{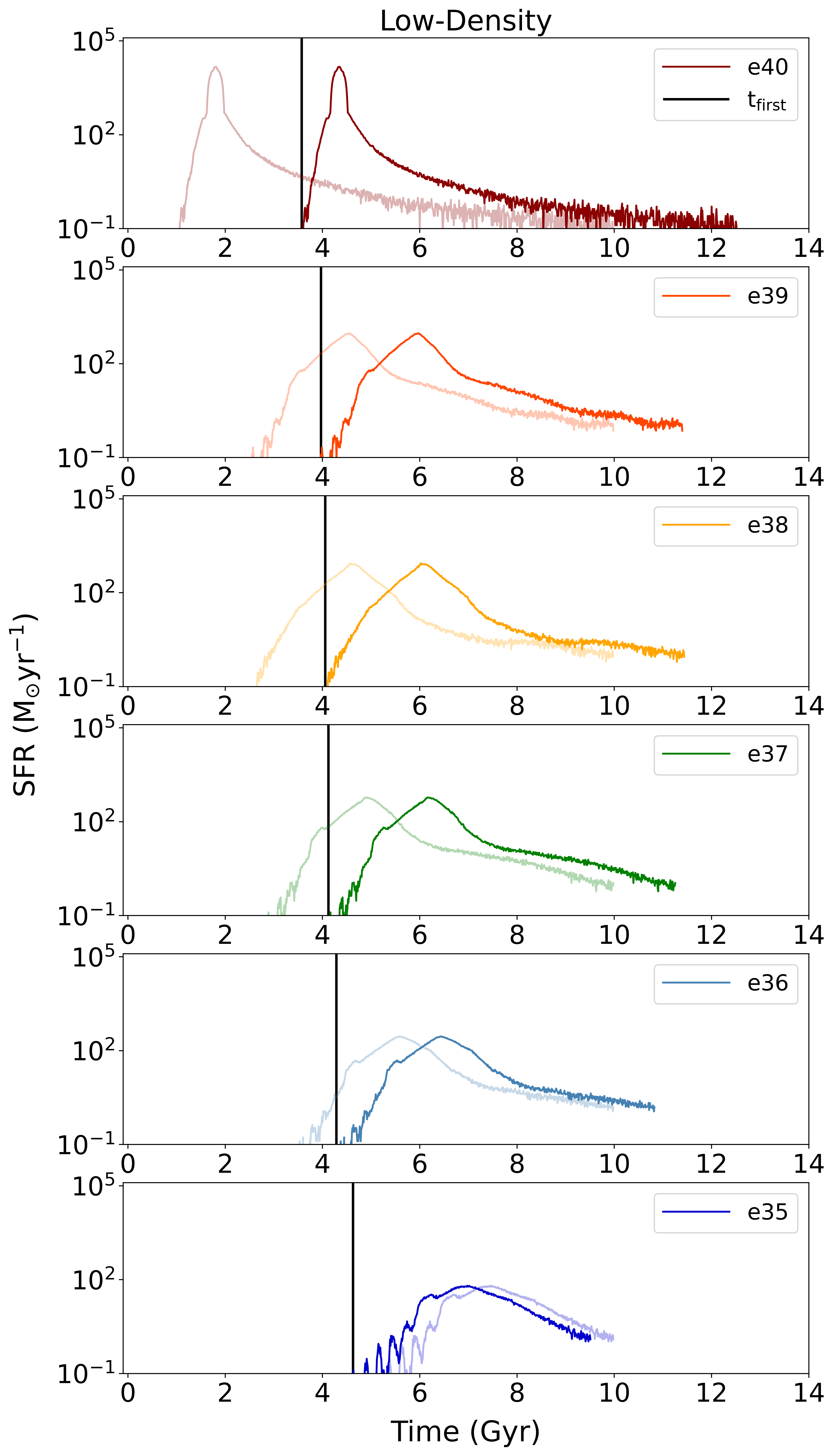}
	\includegraphics[width=\columnwidth]{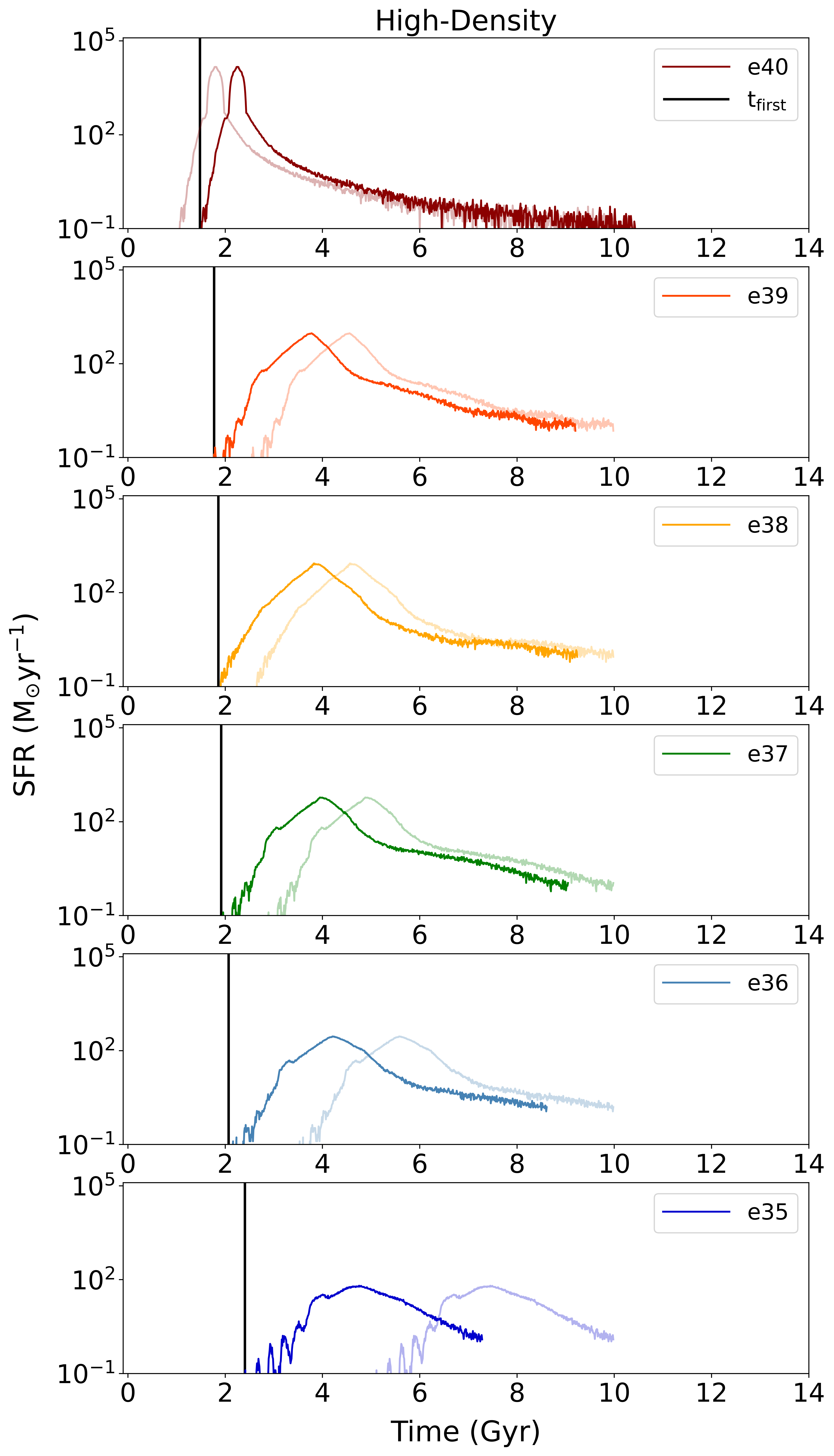}
    \caption{A plot showing the SFH of model e40, e39, e38, e37, e36 and e35 (Table \ref{tab:init_cond}) in the low-density (\emph{left}) and high-density (\emph{right}) environment. The black line is the time when the first star is formed in the models. Each model has been shifted in time according to the \citet{2005ApJ...621..673T} age-dating (Eqs. \ref{eq:tshift} \& \ref{eq:tfirst}). The nearly transparent SFHs are relative to the start of the computation at $t$ = 0.}
    \label{fig:sfhfour}
\end{figure*}

Fig. \ref{fig:bestdown} shows that only models with $R_{\rm initial}$ = 500 kpc ($\Delta\tau_{\rm m}/\Delta\it\tau_{\rm{thomas}}$ $\approx$ 1) are comparable to the results from \citet{2005ApJ...621..673T}. From here, we only consider models with $R_{\rm initial}$ = 500 kpc for discussions. Using the basic assumption that the standard age of the Universe is 13.8 Gyr and the age-dating deduced by \citet{2005ApJ...621..673T} for ETGs is valid, we translate the SFH of our models to a time related to the start of the Big-Bang. Since our models have SFTs comparable to \citet{2005ApJ...621..673T}, we assume that the peaks of the SFHs of our models should coincide with the average ages computed by \citet{2005ApJ...621..673T}. This allows us to place our models to a time relative to the Big-Bang (Eq. \ref{eq:thomasage}). We shift the SFHs of our model galaxies to a time corresponding to the average ages as deduced by \citet{2005ApJ...621..673T}. To achieve this, we first transform the observed average formation times (Eq. \ref{eq:thomasage}) to time since the Big-Bang,
\begin{equation}
    t_{\rm {av,th}}/{\rm Gyr} = 13.8 - t_{\rm{thomas}}/{\rm Gyr},
	\label{eq:tshift}
\end{equation}
and the time the first stellar particles would have formed according to the models presented here becomes,
\begin{equation}
    t_{\rm{first}}/{\rm Gyr} = t_{\rm{av,th}}/{\rm Gyr} - ( t_{\rm{peak}}/{\rm Gyr} - t_{\rm{start}}/{\rm Gyr}),
	\label{eq:tfirst}
\end{equation}
where $t_{\rm{start}}$ is the time when the first stellar particle is formed in the simulation and $t_{\rm{first}}$ is the time when the first star would have formed in the ETG relative to the Big-Bang that is consistent with the \citet{2005ApJ...621..673T} age-mass relation. Fig. \ref{fig:tavvmfinal} shows the average age and when the first star forms in the models for both low- and high-density environments. Fig. \ref{fig:sfhfour} is a representation of the SFH of the model galaxies which follow the downsizing timescales as deduced by \citet{2005ApJ...621..673T}. That is, a massive ($\approx \mathrm{10^{12}} \ M_{\odot}$) post-Big-Bang gas cloud would have started to form an ETG around 1.5 Gyr after the Big-Bang in a high-density environment and 3.6 Gyr after the Big-Bang in a low-density environment. The observed quasar activity only $\approx$ 0.70 Gyr after the Big-Bang suggests that there may have been significantly more over-dense regions \citep{2020MNRAS.498.5652K}. The transparent SFHs in Fig. \ref{fig:sfhfour} are on the time axis where $t$ = 0 is the start of the simulation. Thus, the real Universe appear to have accelerated the start of the collapse even in the low-density regions. Rather than the real Universe having accelerated the onset of star formation, another interpretation of Fig. \ref{fig:sfhfour} is possible though. Fig. \ref{fig:sfhfour} shows that the low-mass models agree with the observed galaxies in terms of the time when the first stars formed in low-density environments, while the massive models agree with the observed timings in high-density regions. This suggests that low-density regions form preferentially low-mass galaxies, while high-density regions form preferentially massive galaxies. This will be testable with cosmological structure formation simulations in MOND.

%\begin{figure}
%	\includegraphics[width=\columnwidth]{figures/lowdennewsfr.png}
%    \caption{A plot showing the SFH of model e41, e39, e38, e37, e36 and e35 (Table \ref{tab:init_cond}) in the low-density environment. The black line is the time when the first star is formed in the models.}
%    \label{fig:sfhfour}
%\end{figure}

%\begin{figure}
%	\includegraphics[width=\columnwidth]{figures/highdennewsfr.png}
%    \caption{Same as Fig. \ref{fig:sfhfour} but in the high-density environment.}
%    \label{fig:sfhfourHD}
%\end{figure}

%\begin{figure}
%	\includegraphics[width=\columnwidth]{figures/taufirstvmass.png}
%    \caption{
%    A plot showing the time, $t_{\rm first}$ vs $M_{\rm final}$ for the models listed in Table \ref{tab:init_cond} with $R_{\rm initial}$ = 500 kpc in both high-density (HD) and low-density (LD) environment.}
%    \label{fig:taufirst}
%\end{figure}
%%%%%%%%%%%%%%%%%%%%%%%%%%%%%%%%%%%%

\subsection{Effective-radii of the model galaxies}
\label{sec:effrad}
\begin{figure}
	\includegraphics[width=\columnwidth]{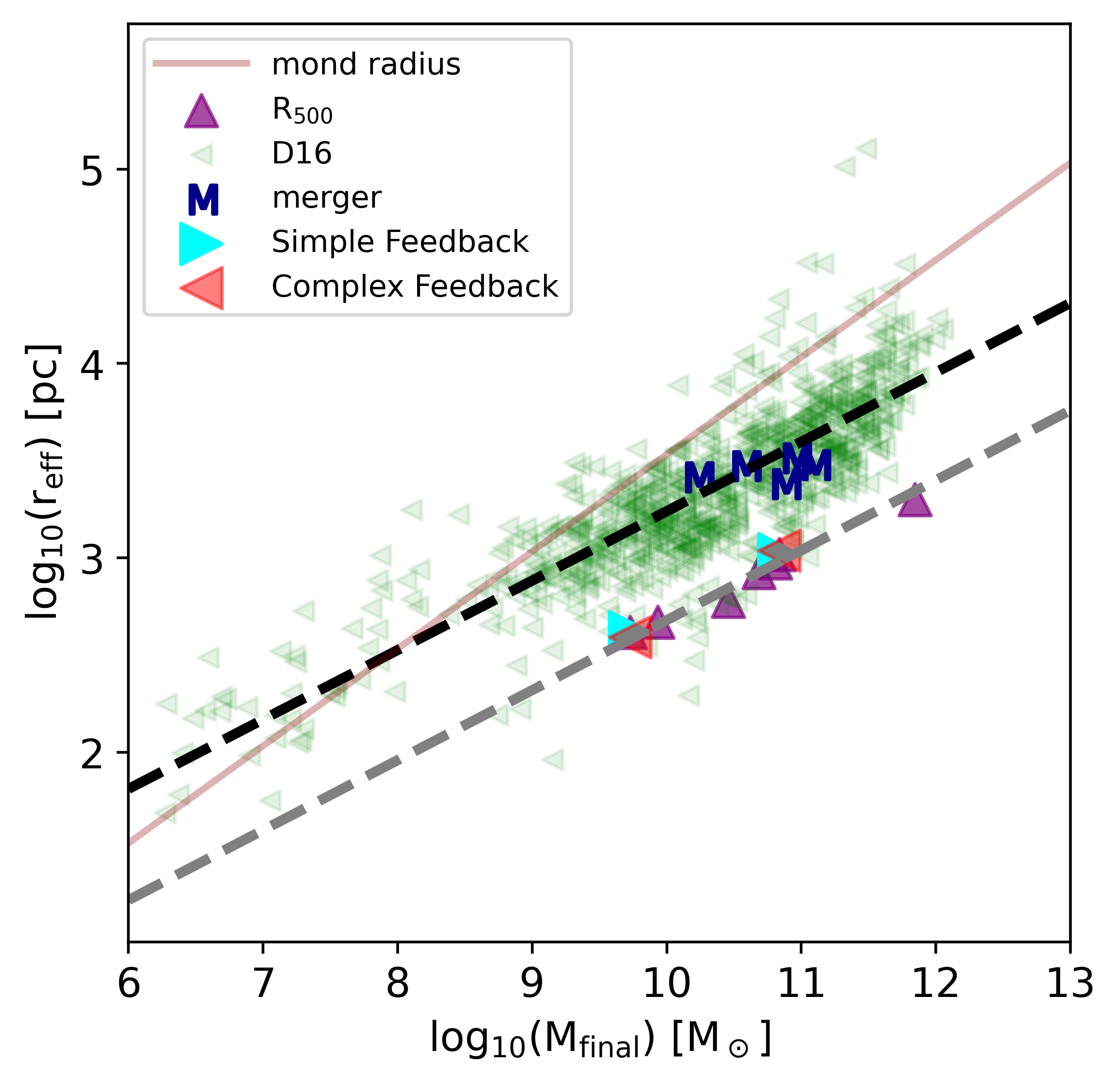}
    \caption{The effective-radius--mass relation. The green triangles are the real ETGs from \citet{2016MNRAS.460.4492D}, purple triangles are the model galaxies e34 to e40 with $R_{\rm initial}$ = 500 kpc listed in Table \ref{tab:init_cond} and the blue M's are the merged models listed in Table \ref{tab:merger}. The black dashed line is the best fit for the real galaxies (Eq. \ref{eq:reffdabri}) and the grey dashed line is the best fit for the model galaxies (Eq. \ref{eq:reffmodels}). The solid red line is the MOND radius (Eq. \ref{eq:mondradius}) and the red left-pointing triangles are the models with complex feedback (see Section \ref{sec:complexfeedback}).}
    \label{fig:reffvmf}
\end{figure}

The effective-radius is defined as the projected radius within which half of the final stellar mass of the model galaxy is found, assuming that the mass density follows the luminosity density, this can be compared to the projected half-light radius from observations. The effective-radius is calculated by projecting the model galaxy into the XY plane (Eappen et al. 2022c will further discuss the effective-radius in other projections as well). The shape of the model galaxy is found to be disk-like similar to \citet{2020ApJ...890..173W}. The collapse occurs along the $Z$-direction such that the mid-plane of the formed galaxy is the XY plane. The effective-radius, $r_{\rm eff}$, is plotted against $M_{\rm final}$ of the model galaxies and compared to the real galaxies from \citet{2016MNRAS.460.4492D} in Fig. \ref{fig:reffvmf}. 

The MOND radius, $r_{\rm M}$, at which the Newtonian radial acceleration equals $a_{\rm0}$ \citep{2014SchpJ...931410M} is,

\begin{equation}
    r_{\rm {M}} = \left(\frac{\mathrm{G} \cdot M_{\rm final}}{a_{0}}\right)^{0.5} ,\\
	\label{eq:mondradius}
\end{equation}
and is plotted as the solid red line in Fig. \ref{fig:reffvmf}. \citet{2008MNRAS.386.1588S} notes that the effective radius of the dissipationless models in his work are comparable to $r_{\rm M}$ that is, galaxy mass ($\approx 10^{11} \ M_{\odot}$) objects naturally collapse to a radius of about 10 kpc. The hydro-dynamical dissipative simulations of the models in this work form model galaxies similar in shape, size and kinematical properties to the compact relic galaxies like NGC 1277 \citep{2014ApJ...780L..20T} and this will be discussed in a companion paper (Eappen et al. 2022b in preparation).

The best-fit relation for the \citet{2016MNRAS.460.4492D} galaxies in Fig. \ref{fig:reffvmf} is given by,
\begin{equation}
    %r_{\rm {eff}}/{\rm pc} = 0.15 \cdot {\rm log_{10}}(M_{\rm final}/M_{\odot}) + 1.22 \\
	r_{\rm {D}}/{\rm pc} = 0.46 \cdot (M_{\rm \star}/M_{\odot})^{0.35},
	\label{eq:reffdabri}
\end{equation}
where $r_{\rm D}$ is the observed effective-radius calculated by \citet{2016MNRAS.460.4492D} and $M_{\star}$ is the total stellar mass of the galaxy. The best-fit for the model galaxies in Fig. \ref{fig:reffvmf} is calculated using a power law function,
\begin{equation}
    %r_{\rm {eff}}/{\rm pc} = 0.15 \cdot {\rm log_{10}}(M_{\rm final}/M_{\odot}) + 1.22 \\
	r_{\rm {eff}}/{\rm pc} = 0.12 \cdot (M_{\rm final}/M_{\odot})^{0.36}.
	\label{eq:reffmodels}
\end{equation}
It can be seen in Fig. \ref{fig:reffvmf} that the effective-radius -- mass relation (Eq. \ref{eq:reffmodels}) of the model galaxies has a power-law index comparable to the effective-radius -- mass relation (Eq. \ref{eq:reffdabri}) of the observed ETGs from \citet{2016MNRAS.460.4492D} but that it falls below the observed relation by a factor of 4.

It is possible that the monolithic collapse which forms individual compact ETGs occur in the inner regions of local density enhancements that form neighbouring gas clouds which collapse monolithically, forming binary or even more ETGs (similar to collapse of stars in a very dense configuration e.g. \citealt{1985MNRAS.214...87J}, \citealt{1988MNRAS.233....1W} and \citealt{2014ApJ...789..115S}). If these are on radial orbits they would merge, leaving a merger remnant which is likely to have a larger effective-radius (For further empirical evidence that early-type galaxies formed first at their innermost regions see section 2.2 in \citealt{2020MNRAS.498.5652K}). \citet{2007MNRAS.382..109T} study an effective size evolution of compact galaxies where they find that the massive ETGs were formed in the early Universe, and have subsequently evolved passively until today. They find an effective size evolutionary mechanism would be able to evolve their compact galaxies to the observed local relation with just two major mergers. To investigate this we study the merger scenario where two model galaxies listed in Table \ref{tab:init_cond} are placed 100 kpc apart in a box of 1 Mpc. The models only consist of stellar particles. The spin angular momentum of each model galaxies is retained as they are placed in the new box and they merge under MOND gravity to form a new merged model galaxy in 1.5 Gyr. Table \ref{tab:merger} lists the initial conditions and results of the merged models. Five different merger studies were done placing the models at different axes and at different orientation. M1, M2, M3 and M4 are merged models formed from the merger of the same model galaxies and M5 is formed by the merger of model galaxies of different masses (e39 and e36) that merge at different angles with respect to the XY plane (where parent models e36 and e39 are placed tilted at an angle of 45 degrees with respect to the XY plane). The effective-radius--mass relation of these merged models is plotted in Fig. \ref{fig:reffvmf}. It can be seen that the merged models lie around the same dashed black line as the real ETGs in Fig. \ref{fig:reffvmf}. A detailed study of the morphology and other structural properties of these merged galaxies will be discussed in a follow up paper (Eappen et al. 2022c).

Another possible mechanism for the model galaxies to reach large radii is through stellar evolution mass loss which is directly linked to the galaxy-wide IMF assumed. It has been shown that the stellar remnant population is comparable to the mass of living stars in the IGIMF theory but using a canonical invariant IMF would decrease the remnant mass to be much lower than the stellar mass \citep{2019A&A...629A..93Y}. \citet{2021A&A...655A..19Y} finds that the ratio of dynamical mass and stellar mass, $M_{\rm dyn}/M_{\star}$, is higher for more massive galaxies because they have a more top-heavy galaxy-wide IMF and more stellar remnants. Given the high SFRs (> $10^{2}M_{\rm \odot}/\rm yr$) the model galaxies reach (Table \ref{tab:init_cond}), it is expected that the IGIMF is top-heavy \citep{2017A&A...607A.126Y,2018A&A...620A..39J,2021A&A...655A..19Y} leading to significant mass loss from the evolving stars. This mass lost would be in the form of gas and if it is heated sufficiently enough it would expand and lead to the expansion of the galaxy \citep{2022arXiv220614238S}.
%The integrated galaxy-wide IMF (IGIMF), is a theory to calculate the IMF of the whole galaxy by summing all the IMF's of all embedded clusters that form in the galaxy \citep{2003ApJ...598.1076K, 2006MNRAS.365.1333W} and the shape of this galaxy-wide IMF is described, depending on the metallicity and SFR, as top-light or top-heavy and bottom-light or bottom-heavy \citep{2017A&A...607A.126Y,2018A&A...620A..39J,2021A&A...655A..19Y}. The model galaxies here assume an invariant IMF, the models do not expand to reach the $r_{\rm eff}$ of observed galaxies. A possible way this might naturally be solved is if the expansion is based on a variable galaxy-wide IMF \citep{2009MNRAS.394.1529D, 2009ApJ...706..599L, 2012Natur.484..485C, 2012ApJ...747...72D, 2018Natur.558..260Z}. The integrated galaxy-wide IMF (IGIMF), is a theory to calculate the IMF of the whole galaxy by summing all the IMF's of all embedded clusters that form in the galaxy \citep{2003ApJ...598.1076K, 2006MNRAS.365.1333W} and the shape of this galaxy-wide IMF is described, depending on the metallicity and SFR, as top-light or top-heavy and bottom-light or bottom-heavy \citep{2017A&A...607A.126Y,2018A&A...620A..39J,2021A&A...655A..19Y}.  This IGIMF theory can be incorporated to calculate how the collpase models would expand radially which will be investigated in the future.
%model re mass
%70M70M 3000 1.36E11
%10M10M 2600 0.17E11
%rot30M70M 2400 0.85E11
%30M30M 2980 0.38E11
%50M50M 3250 0.9E11
\begin{table}
\caption{Initial conditions of the merger models}
\label{tab:merger}
\begin{tabular}{c c c c}
\hline
Model & Parent & $r_{\rm eff}$ & $M_{\rm final}$\\
Name & models & (kpc) & ($M_{\rm \odot}$)\\
\hline
M1 & e35 + e35 & 2.6 & 0.17E11\\
M2 & e36 + e36 & 2.9 & 0.38E11\\
M3 & e37 + e37 & 3.2 & 0.9E11\\
M4 & e39 + e39 & 3.0 & 1.25E11\\
M5 & Re36 + Re39 & 2.4 & 0.75E11\\
\hline
\end{tabular}
\\Note: Column 1: name, Column 2: The models from Table \ref{tab:init_cond} that merge, Column 3 ($r_{\rm eff}$): the effective-radius of the merged model and Column 4 ($M_{\rm final}$): the mass of the merged model. Models Re36 and Re39  are models e36 and e39 but rotated (see text for details).
\end{table}

%%%%%%%%%%%%%%%%%%%%%%%%%%%%%%%%%%%%%%%%%%%%%%%
\subsection{Effects of complex feedback mechanism}
\label{sec:complexfeedback}

\begin{figure*}
	\includegraphics[width=\columnwidth]{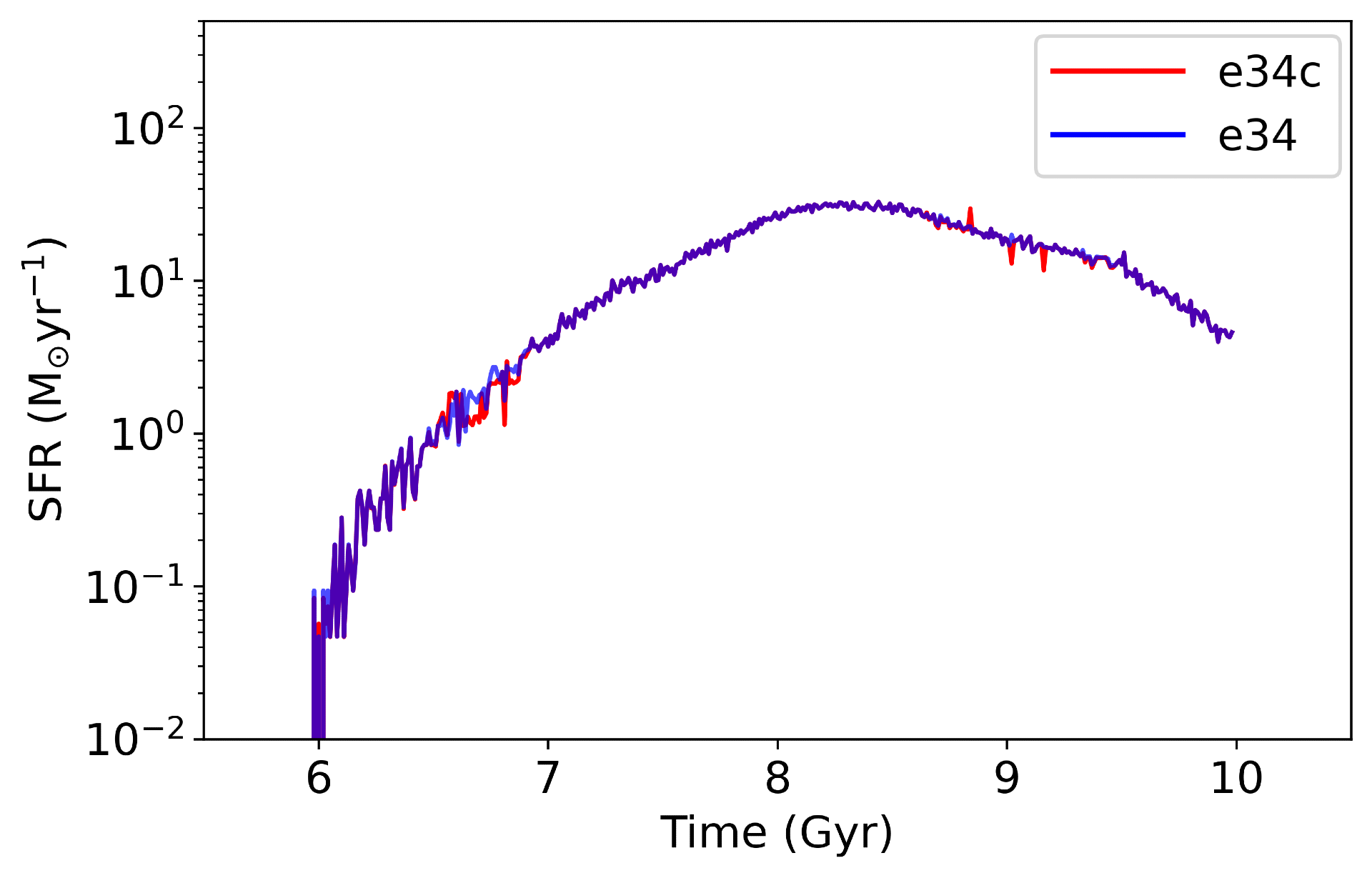}
	\includegraphics[width=\columnwidth]{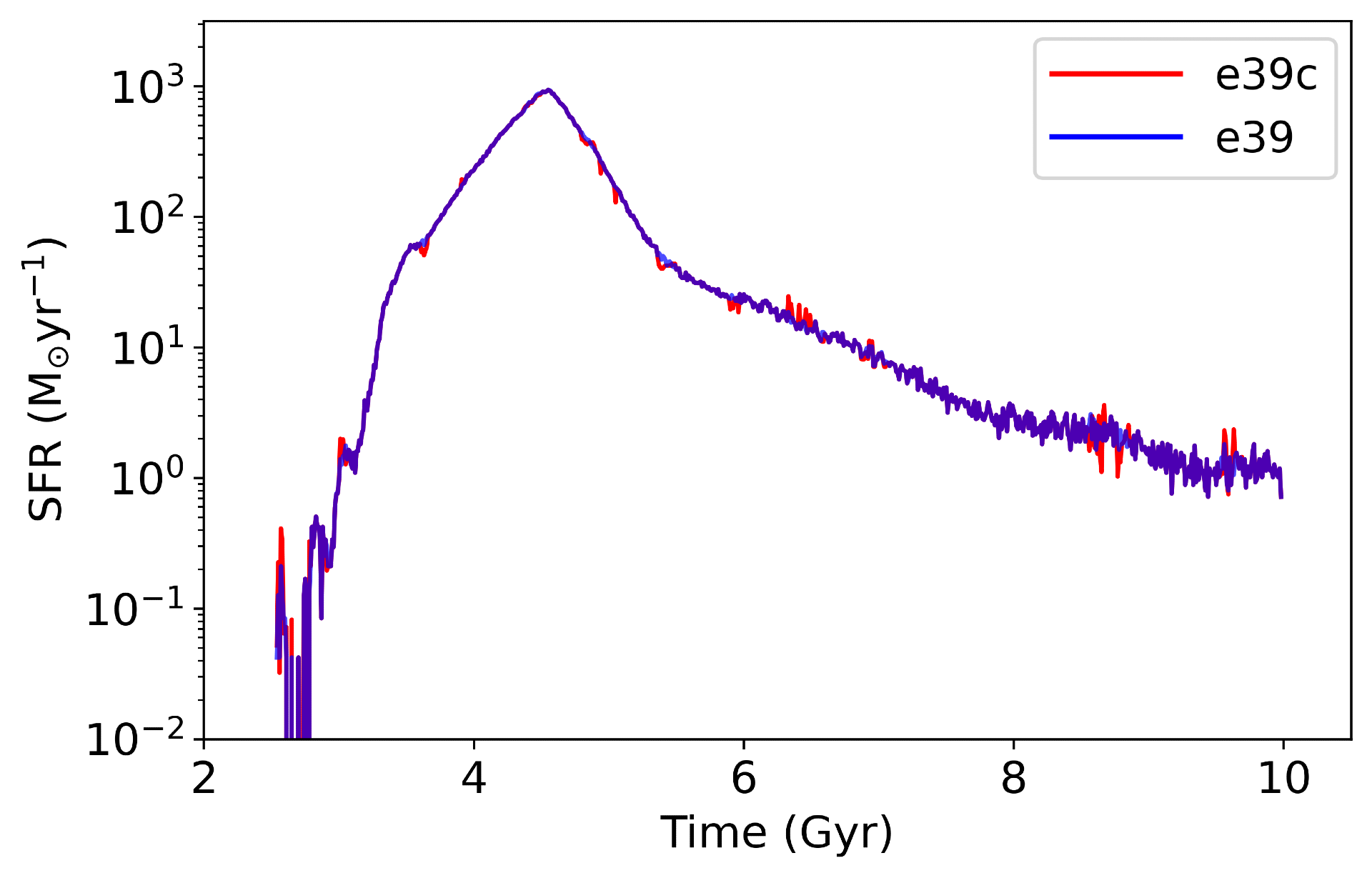}
    \caption{A plot comparing the SFH of models with simple and complex feedback. Models e34c and e39c are models with complex feedback.}
    \label{fig:SFCF}
\end{figure*}

The above results (except for models e34c and e39c) have been obtained using the simple feedback algorithm (simple cooling/heating and no additional more complex baryonic physics, such as supernovae). Models e34c and e39c were computed with additional feedback process to investigate how complex baryonic physics affects the results obtained.

The simple star-formation algorithm only considers the heating/cooling modules in PoR. Cooling and heating of the gas are computed by using tables that are included in the code, which describe the cooling and heating rates due to several physical processes. PoR uses \citet{2004A&A...416..875C} look-up tables for the different cooling/heating processes. The complex feedback algorithm includes more initial inputs such as supernovae, radiative transfer, sink particles and a higher refinement level which increases the resolution  of the simulation. The supernova (and in fact all feedback scheme) scheme is implemented as in RAMSES where the kinetic part of the supernova energy is injected as a spherical blast wave with the size of galactic superbubbles of radius 150 pc. The radiative transfer option allows to compute the radiative transfer between stellar particles with the addition of photon fluxes which uses a first-order Godunov solver \citep{2013MNRAS.436.2188R}. A sink particle scheme is used to stabilize the simulation by artificially stopping the collapsing gas cloud if a certain density is exceeded and the gas is condensed into a point mass (sink particle, see \citealt{2002A&A...385..337T, 2015CaJPh..93..232L, 2020ApJ...890..173W,2021CaJPh..99..607N} for a detailed description on how RAMSES and PoR handles feedback).

\begin{table}
\caption{Effects of feedback}
\label{tab:feedback}
\begin{tabular}{c c c c}
\hline
 & Simple Feedback & Complex Feedback & \\
\hline
 & e34 & e34c & PC\\
$r_{\rm eff}$ & 0.41 & 0.39 & - 4.87\%\\
$\Delta\tau_{\rm m}$ & 1.41 & 1.43 & + 1.41\%\\
\hline
 & e39 & e39c & PC\\
$r_{\rm eff}$ & 1.04 & 1.09 & + 4.80\%\\
$\Delta\tau_{\rm m}$ & 0.54 & 0.53 & - 1.85\%\\
\hline
\end{tabular}
\\Note: PC shows the percentage change (Eq. \ref{eq:percentchange}) in the results obtained from models with same $R_{\rm initial}$ and $M_{\rm initial}$ but with simple (e34 $\&$ e39) and complex (e34c $\&$ e39c) feedback.
\end{table}

The model e39c has the same $R_{\rm initial}$ and $M_{\rm initial}$ as model e39 but is computed with complex baryonic feedback of supernovae, sink particles, and radiative transfer with a higher resolution (maximum refinement level set to 13 instead of 12). Similarly model e34c has the same initial conditions as model e34 but is computed with a higher resolution (maximum refinement level set to 14 instead of 12). Fig. \ref{fig:SFCF} shows the SFH of the models with simple and complex feedback. Table \ref{tab:feedback} shows the effect of feedback on the models which have the same $R_{\rm initial}$ and $M_{\rm initial}$. The percentage change (PC) is calculated using Eq. \ref{eq:percentchange} in Appendix \ref{sec:percentchange} and the results obtained (Table \ref{tab:feedback}) show that the percentage change of the results from the complex feedback is within 5$\%$ of the results obtained from simple feedback. The complex feedback is significantly more CPU intensive but is found to affect the results negligibly (see Fig. \ref{fig:downsizing}, \ref{fig:reffvmf} and \ref{fig:SFCF} and Table \ref{tab:feedback}), as already shown in the disk-galaxy formation simulations by \citet{2020ApJ...890..173W}. %The model e40 is the same as the model e39 but with complex baryonic feedback of supernovae, sink particles, and radiative transfer with a higher resolution (maximum refinement level set to 13 instead of 12). It can be seen in Fig. \ref{fig:downsizing} and Fig. \ref{fig:reffvmf} that model e40 does not differ in any significant way from model e39.
This demonstrates again that differences in the feedback processes play a minor role in the evolution of galaxies and that the physics of galaxy evolution are pre-dominantly defined by the Milgromian law of gravitation \citep{2015CaJPh..93..169K}.

%%%%%%%%%%%%%%%%%%%%%%%%%%%%%%%%%%%%%%%%%%%%%%%%%%
%%%%%%%%%%%%%%%%%%%%%%%%%%%%%%%%%%%%%%%%%%%%%%%%%%

%%%%%%%%%%%%%%%%%%%%%
\section{Discussion}
\label{sec:discussion}
%on comparing to the real Universe.
The monolithic post-Big-Bang collapse of non-rotating gas clouds in MOND results in a SFT--mass relation (Eq. \ref{eq:50kpc}) which is comparable to the observed relation for ETGs \citep{2005ApJ...621..673T,2010MNRAS.404.1775T}. Collapsing post-Big-Bang gas clouds in Milgromian gravitation thus behave as the observed ETGs in terms of the collapse timing (i.e more massive clouds collapse faster and form their stellar particles more quickly as shown in Fig. \ref{fig:denvtaumodels}). The grid of models with different $R_{\rm initial}$ and $M_{\rm initial}$ allows us to constrain $R_{\rm initial} \approx$ 500 kpc as best fitting the observed SFT--mass relation. This suggests that typically a region spanning almost a Mpc across collapsed to form ETGs if sufficiently massive. We note that field elliptical galaxies are slightly younger (consistent with the results found by \citealt{2005ApJ...621..673T}) but otherwise do not differ significantly from elliptical galaxies in clusters \citep{2017A&A...597A.122S}. This suggests that also in the field the bulk stellar population of elliptical galaxies formed through monolithic collapse on the downsizing time scale.

How do these results compare to the expectations from the standard hierarchical structure formation theory as quantified through the Illustris TNG \citep{Pillepich_2018,2018MNRAS.475..624N,Nelson_2019} and Eagle \citep{Schaye_2015,McAlpine_2016} projects? It is noted that \citet{2010MNRAS.405..705F} write: \textit{"The so-called downsizing trend in galaxy formation (e.g. more massive galaxies forming on a shorter time-scale and at higher redshift with respect to lower mass counterparts) is not fully recovered."} On the other hand, based on their analysis of the Millenium Simulation, \citet{2006MNRAS.366..499D} claim \textit{"These findings are consistent with recent observational results that suggest ‘down-sizing’ or ‘anti-hierarchical’ behaviour for the star formation history of the elliptical galaxy population, despite the fact that our model includes all the standard elements of hierarchical galaxy formation and is implemented on the standard, $\Lambda$CDM cosmogony."} Given these discrepant results based on the same theory and observational data, we evaluate the formation time scales of early type galaxies that form in the Illustris TNG and EAGLE projects (see Figs.~\ref{fig:tngtau} and \ref{fig:lcdmsft2}), noting that both of these rest on very different simulation codes (adaptive mesh refinement versus SPH, respectively) and also very different baryonic physics algorithms. The details are provided in Appendix \ref{sec:lcdm simulations} and the result is, in summary, that the formation time scales of early type galaxies formed in both of these simulation projects do not concur with the observed population.

%The SFH of the model galaxies in our simulations were shifted to the average ages of real galaxies to calculate the time when the first stars would have formed in ETGs. This is done under the assumption that the standard age of the Universe is 13.8 Gyr and the average age of ETGs are as deduced by \citet{2005ApJ...621..673T}. This result follows the downsizing and allows us to estimate the time when the first stars are formed in the ETGs since the Big-Bang in both the high-density and low-density environments. We find that our massive model has a very high SFR at the peak (Fig. \ref{fig:sfhplot}), forms the first stellar particle in a short timescale and soon ($\approx$ 1 Gyr) after the Big-Bang (Fig. \ref{fig:taufirst}). This is very similar to the recent observations from \citet{2018Natur.553...51M} who report a massive galaxy at a redshift of 6.9 that would have formed less than 1 Gyr after the Big-Bang and which has a SFR of 2900 $M_{\odot}/\mathrm{yr}$. As just shown, the age-mass scaling by \citet{2005ApJ...621..673T} in high density regions leads to the first stars forming at about 1 Gyr after the Big Bang, which is later than needed to account for the observed extremely high-redshift quasars. This problem is discussed in \citet{2020MNRAS.498.5652K} who need to introduce an additional dispersion of densities such that the first star formation in the highest density peak would take place about 100 Myr after the nominal Big-Bang in order to account for the quasars and SMBHs in ETGs (Fig. 15 in \citealt{2020MNRAS.498.5652K}). 

The MOND radius (Eq. \ref{eq:mondradius}) proposed by theory is larger but follows the same trend as the observed $r_{\rm eff}$. This similarity of the MOND radius and the effective-radius was first noted by \citet{2008MNRAS.386.1588S} where he concludes that a dissipationless initially expanding gas cloud collapse would naturally form galaxies with effective-radii that are comparable to their MOND radii. The models in our study do not consider an expanding gas cloud collapse and fall short of this expected effective-radius--mass relation (see Fig. \ref{fig:reffvmf}) but this is possibly also due to a non-variant  galaxy-wide IMF which makes the post-Big-Bang clouds to collapse too deeply. If, on the other hand, the galaxy-wide IMF varies such that it becomes top-heavy at a high SFR (eg. \citealt{2018A&A...620A..39J,2021A&A...655A..19Y}) then the collapse is likely less deep and $r_{\rm eff}$ will end up larger. We do not address the chemical evolution of the models here as the chemical enrichment has not yet been incorporated into PoR and would also require a systematically varying galaxy-wide IMF (the current simulationss effectively assume the canonical IMF for the formation of stellar particles). The SFHs of the models presented here are however comparable, by virtue of them matching the formation timescales, with the constraints from abundance and stellar population observations as analysed by \citet{2021A&A...655A..19Y}. It is to be expected that the bulk chemical enrichment of the models computed here would follow the closed box models of \citet{2021A&A...655A..19Y}. The problem of spatially-resolved chemical enrichment of the forming model ETGs will be studied in the future in connection with the systematically varying galaxy-wide IMF.

The model galaxies in this work do not agree with the SFT--mass relation found in \citet{2021A&A...655A..19Y} (dotted green curve in Fig. \ref{fig:downsizing}). This is possibly due to the non-variant IMF that is assumed in the simulation as mentioned above. The other method through which the model galaxies could reach the \citet{2021A&A...655A..19Y} SFT--mass relation is if the gas cloud collapse calculations are considered in an expanding cosmological volume which will be investigated in the future.

In Fig. \ref{fig:reffvmf} we show that mergers of the model galaxies form models that lie on the observationally deduced effective-radius--mass relation. These kinds of mergers could happen in the initially densest parts of the Universe, for e.g. in central regions of galaxy clusters, where the galaxies would have initially formed through monolithic collapse of post-Big-Bang gas clouds and then slightly later merge with another galaxy of comparable mass in a timescale somewhat larger than the formation timescale of these galaxies to reach the observed effective-radius--mass relation. The morphology and structural properties of the model galaxies and the mergers using different projections will be discussed in a follow up paper (Eappen et al. 2022c in preparation) where we find that the shape of these merged models is triaxial and very similar to the observed ETGs. Thus, the formation timescales and other observed chemical properties of ETGs could have been embedded in them before they undergo one or two major mergers such that only the structural properties and the morphology of these ETGs changed through the mergers. \citet{2007MNRAS.382..109T} reports a similar result in their study of size evolution of massive galaxies at different redshifts. They find that at z $\approx$ 1.5 massive spheroid like objects were a factor of 4 smaller than we see today and suggest that just two major mergers of equal mass would evolve the size of the galaxy to what we observe today. Similarly, \citet{2008A&A...486..763P} in their study of ETGs through chemical evolution models find that a series of dry mergers of galaxies is not the way to recover downsizing and suggest 1--3 major-dry mergers to be in agreement with the observations. Our results show that it is possible to evolve the size and mass of the galaxy without changing the intrinsic properties of the galaxy such as the SFT without a large number of mergers.

%As shown in Section \ref{sec:complexfeedback}, the complex feedback mechanism does not affect the results in this work significantly. 
%\textcolor{red}{Maybe mention \citet{2021arXiv210813430C}}

\section{Conclusion}
\label{sec:Conclusion}
In this work, we discuss the formation of ETGs in comparison with observational results. The isolated monolithic collapse of post-Big-Bang clouds of different initial radii at fixed mass is computed in Milgromian gravitation and is found to produce model galaxies with SFTs similar to the observed ETGs.

The main results are:
\begin{itemize}
    \item This work for the first time shows that the SFTs observed for ETGs is a natural occurrence in the MOND paradigm if ETGs form from monolithically collapsing non-rotating post-Big-Bang gas clouds. We were able to reproduce the SFTs as noted by observations (Fig. \ref{fig:downsizing}). The models with $R_{\rm initial}$ = 500 kpc are similar to the relation of \citet{2005ApJ...621..673T,2010MNRAS.404.1775T} and the models with $R_{\rm initial}$ = 200 kpc fit fairly closely to the relation of \citet{2009A&A...499..711R}. This work thereby gives a rough estimate of the initial conditions of the post-Big-Bang clouds in order to produce ETGs. The results indicate that the observed Universe started to form ETGs earlier than the here studied collapse models, although the SFTs of the models match those observed for gas clouds with initial radii in the range 200 to 500 kpc. Higher resolution simulations will be needed to address this issue.
    %\item We find the time $t_{\rm first}$ required to form the first stars in the ETGs since the Big-Bang for a given mass by assuming the standard age of the Universe to be 13.8 Gyr and the age-dating obtained by \citet{2005ApJ...621..673T} is valid. Short $t_{\rm first}$ values are obtained for larger $M_{\rm final}$ as shown in Fig. \ref{fig:taufirst} that is, massive ETGs form their first stellar particles earlier than the less massive ETGs.
    \item The effective-radius--mass relation of the model galaxies falls short of the expected value but the models with only one major merger (Fig. \ref{fig:reffvmf}) are found to be consistent with observations \citep{2016MNRAS.460.4492D}.
    \item We find that different feedback processes affect the results negligibly (as already shown by \citealt{2020ApJ...890..173W} in their formation of disk-galaxies) suggesting that the physics of galaxy formation is pre-dominantly defined by the Milgromian law of gravitation \citep{2015CaJPh..93..169K}.
\end{itemize}
The future aspects of this project will analyze the structural properties and shapes of both the model galaxy and merged galaxies using different projections to calculate the sizes (Eappen et al. 2022c, in preparation). Then the next step would be to investigate how a possible inclusion of a systematically varying galaxy-wide IMF \citep{2017A&A...607A.126Y} would lead to the expansion of the models also taking into account the cosmological expansion of space.
%Notes: \citet{2005MNRAS.362...41G} young metal-poor stellar populations are found predominently in low-mass galaxies. \citet{2020MNRAS.497.3011L}
%%%%%%%%%%%%%%%%%%%%%%%%%%%%%%%%%%%%%%%%%%%%%%%%%%
\section*{Acknowledgements}
\label{sec:acknowledgements}

We would like to thank Zhiqiang Yan for providing insightful suggestions. The numerical simulations were performed using the Tiger--Cluster of the Astronomical Institute of Charles University, Prague. R.E. is supported by the Grant Agency of Charles University under grant No. 234122. B.F. acknowledges funding from the Agence Nationale de la Recherche (ANR projects ANR-18-CE31-0006 and ANR-19-CE31-0017), and from the European Research Council (ERC) under the European Union’s Horizon 2020 Framework programme (grant agreement number 834148). We thank the DAAD Eastern European grant at Bonn University for supporting the research visits.
%%%%%%%%%%%%%%%%%%%%%%%%%%%%%%%%%%%%%%%%%%%%%%%%%%
\section*{Data Availability}
All data used here have been generated as described using the publicly available POR code and the cited literature.
\bibliographystyle{mnras}
\bibliography{example}
% if your bibtex file is called example.bib
%\begin{thebibliography}{99}
%\end{thebibliography}
%%%%%%%%%%%%%%%%%%%%%%%%%%%%%%%%%%%%%%%%%%%%%%%%%%
%%%%%%%%%%%%%%%%% APPENDICES %%%%%%%%%%%%%%%%%%%%%
%\clearpage
\appendix
\section{Fit Parameters}

The fit parameters for Eq.\ref{eq:50kpc} are provided in Table \ref{tab:downtable1}.

\begin{table}[H]
\caption{Fit parameters for Eq.\ref{eq:50kpc}}
\label{tab:downtable1}
\begin{tabular}{c c c}
\hline
$R_{\rm initial}$ & $A$ & $B$\\
(kpc) &  & \\
\hline
50 & 40747.40 & -0.55 \\
100 & 6102.39 & -0.43 \\
200 & 2989.50 & -0.37 \\
300 & 227.87 & -0.25 \\
500 & 6253.16 & -0.37 \\
\hline
\end{tabular}
\end{table}

\section{SFTs for $\Lambda$CDM simulations}
\label{sec:lcdm simulations}
Using the TNG50-1 and TNG100-1 simulation of the Illustris The Next Generation (TNG) project \citep{2017MNRAS.465.3291W,2018MNRAS.477.1206N,2018MNRAS.475..676S,Pillepich_2018,2018MNRAS.475..648P,2018MNRAS.475..624N,2018MNRAS.480.5113M,Nelson_2019} and the Ref-L100N1504 simulation (hereafter EAGLE100) of the EAGLE project \citep{2015MNRAS.450.1937C,Schaye_2015,McAlpine_2016}, the star-formation timescales, SFTs, of the formed model galaxies are investigated within the $\Lambda$CDM framework. These projects are sets of different self-consistent cosmological galaxy formation simulations. The TNG project is consistent with the Planck-2015 \citep{Planck_2016_IllustrisTNG} results: $H_{0} = 67.74\,\rm{km\,s^{-1}\,Mpc^{-1}}$, $\Omega_{\mathrm{b},0} = 0.0486$, $\Omega_{\mathrm{m},0} = 0.3089$, $\Omega_{\Lambda,0} = 0.6911$, $\sigma_{8} = 0.8159$, and $n_{\mathrm{s}} = 0.9667$. Here, the TNG50-1 has a box side of 51.7 co-moving Mpc (cMpc) an initial baryonic mass resolution of $m_{\mathrm{b}} = 8.5 \times 10^{4} \,\rm{M_{\odot}}$, and a dark matter particle mass of $m_{\mathrm{dm}} = 4.5 \times 10^{5}\ \rm{M_{\odot}}$. The TNG100-1 simulation has a box side of $110.7$ co-moving Mpc (cMpc), $m_{\mathrm{b}} = 1.4 \times 10^{6} \,\rm{M_{\odot}}$, and $m_{\mathrm{dm}} = 7.5 \times 10^{6}\,\rm{M_{\odot}}$. An overview of the physical and numerical parameters is given in table 1 of \citet{Nelson_2019}. The EAGLE project assumes the Planck-2013 \citep{Planck_2014_EAGLE} measurements with the cosmological parameters being $H_{0} = 67.77\,\rm{km\,s^{-1}\,Mpc^{-1}}$, $\Omega_{\mathrm{b},0} = 0.04825$, $\Omega_{\mathrm{m},0} = 0.307$, $\Omega_{\Lambda,0} = 0.693$, $\sigma_{8} = 0.8288$, and $n_{\mathrm{s}} = 0.9611$ \citep[see also table~1 of][]{Schaye_2015}. The EAGLE100-1 run has a box size of $100$ cMpc, an initial baryonic particle mass of $m_{\mathrm{b}} = 1.81 \times 10^{6} \rm{M_{\odot}}$, and a dark matter particle mass of $m_{\mathrm{dm}} = 9.70 \times 10^{6} \rm{M_{\odot}}$ \citep[table~2 of][]{Schaye_2015}.

We select present-day ($z=0$) subhaloes with a stellar mass $M_{*} > 10^{9}\,\rm{M_{\odot}}$ and a star formation rate of $\it{SFR} < \mathrm{10^{-3}} \,\rm{M_{\odot}\,yr^{-1}}$. In addition, we remove subhaloes with a Subhaloflag value of 0 in the TNG simulation, ensuring that only galaxies or satellites of a cosmological origin are considered. This gives a final sample of $418$ (45 central galaxies), $4384$ (803 central galaxies) and $2916$ (373 central galaxies) galaxies in the TNG50-1, TNG100-1 and EAGLE100 run, respectively. 

\begin{figure*}
	\includegraphics[scale=0.42]{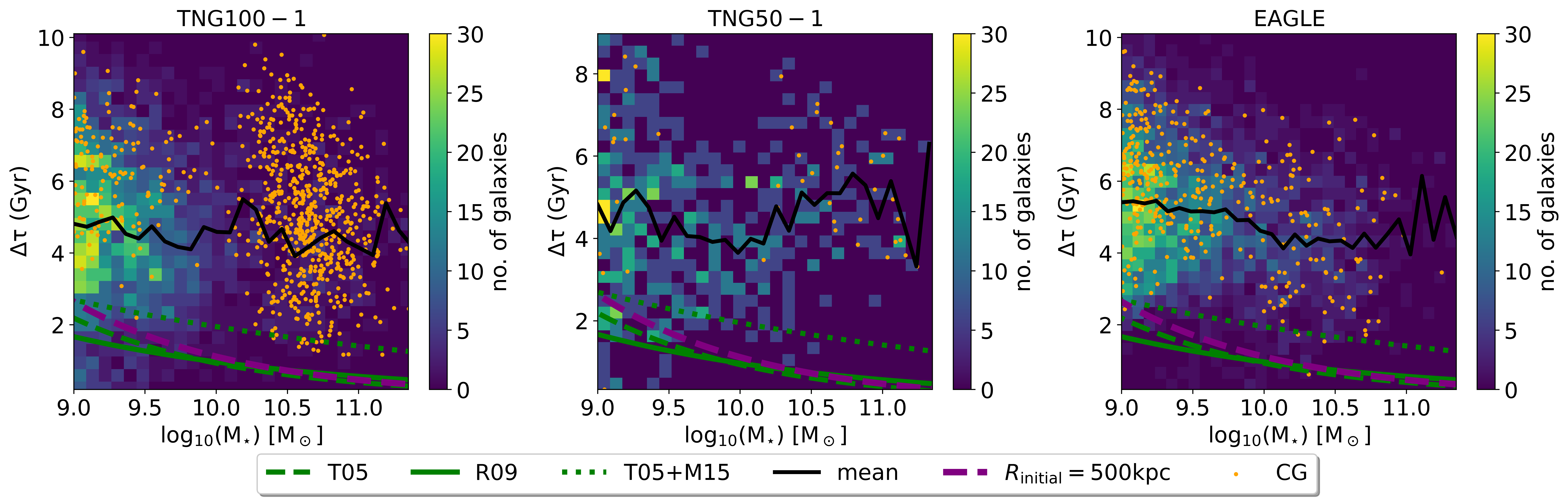}
	\\
	\includegraphics[scale=0.42]{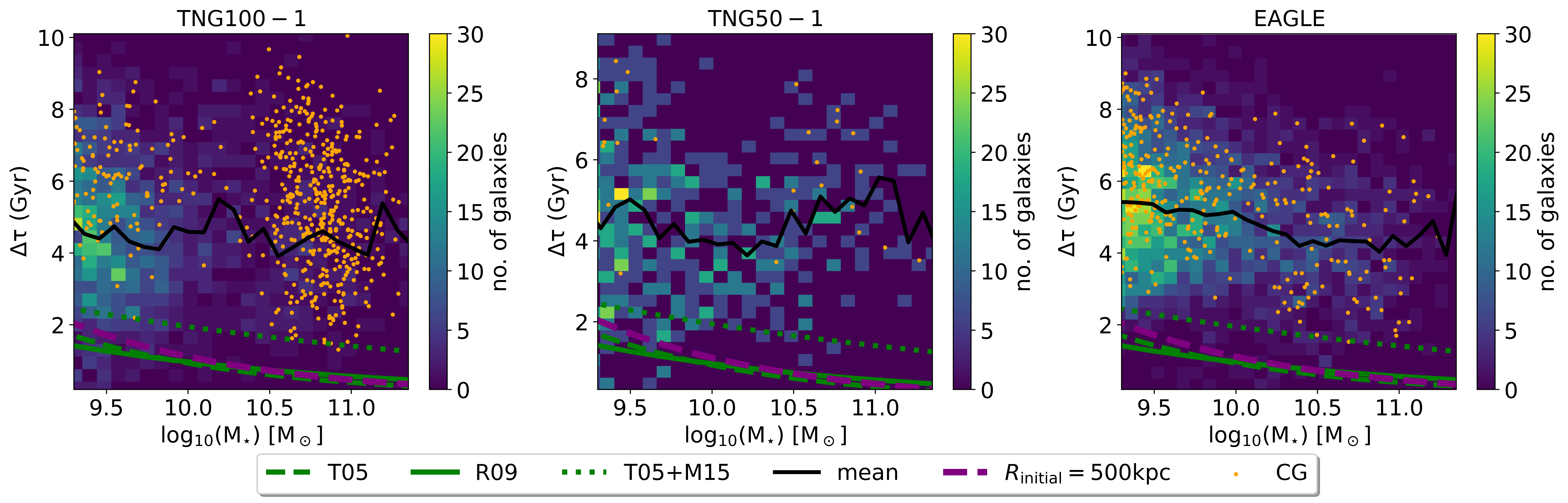}
    %\caption{SFT--mass relation of model galaxies formed in the TNG100-1 (SMoC) simulation compared with observationally constrained relations for ETGs. Note the absence of a trend in the TNG100-1, in stark contrast to observed ETGs.}
    \caption{ %\textcolor{magenta}{Top panel (Method 1): masses of the stellar particles at $z$ = 0, bottom panel (Method 2): initial mass of star particle.} 
    SFT--mass relation of model galaxies formed in the TNG100-1 (\emph{left}), TNG50-1 (\emph{centre}) and EAGLE100 (\emph{right}) (SMoC) simulation by using the method 1 (top panels) and method 2 (bottom panels) compared with observationally constrained relations for ETGs. The orange coloured dots labelled as CG are the central galaxies in each simulation. There are a total of 30 bins on the x- and y-axis and the binsize is 0.08 dex.}

    \label{fig:tngtau}
\end{figure*}

The SFT in the cosmological $\Lambda$CDM simulations is calculated by using three different methods. In the first approach, we extract the formation time and the masses of the stellar particles at $z=0$ of the above selected galaxies. The SFT, $\Delta\tau$, for the TNG100-1 and EAGLE100 data is calculated using the same method as described in Section \ref{sec:models} where the FWHM of the SFH gives the duration for which the bulk of the star formation takes place. The star formation rate (SFR) is calculated by separating all stellar particles in bins of $\delta t$ = 10 Myr according to their age, summing up the stellar mass in every time bin and dividing this by the length of $\delta t$. Their SFT--mass relation is shown in the top panels of Fig. \ref{fig:tngtau}. 

In the second method, we extract again the stellar particles at $z=0$ of the selected galaxies but using the mass of the star particle at the time of its formation in order to calculate the SFT (bottom panels of the Fig. \ref{fig:tngtau}). We use the same method as described above to calculate the SFRs and obtain the SFT--mass relation.

The mean of $\Delta\tau$ for all three simulation runs lies around 5 Gyr and a trend in $M_{*}$ is absent for both methods (except for EAGLE) but is in complete disagreement to observed ETGs \citep{2005ApJ...621..673T,2009A&A...499..711R,2015MNRAS.448.3484M}. The Wilcoxon Signed-Rank test \citep{Bhattacharya} shows the SMoC models to disagree with the observed SFTs at more than 5 sigma confidence because the latter lie consistently and significantly below the former. 

\begin{figure*}
	\includegraphics[scale=0.42]{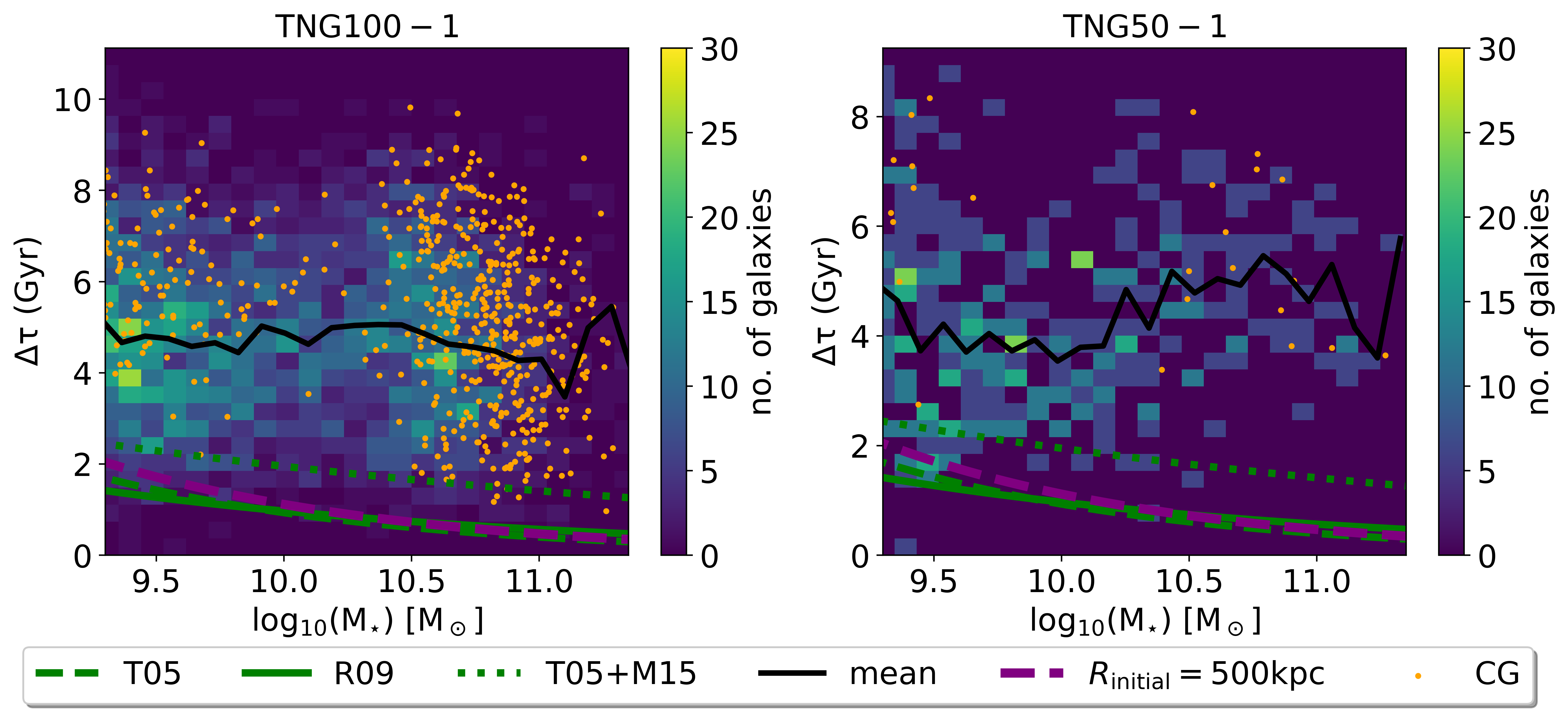}
    %\caption{SFT--mass relation of model galaxies formed in the TNG100-1 (SMoC) simulation compared with observationally constrained relations for ETGs. Note the absence of a trend in the TNG100-1, in stark contrast to observed ETGs.}
    \caption{Same as Fig. \ref{fig:tngtau} but using method 3 as described in Appendix \ref{sec:lcdm simulations}.}

    \label{fig:lcdmsft2}
\end{figure*}

In the third method, we trace the selected galaxies back in cosmic time by using the merger tree catalogs \citep{RodriguezGomez_2015} and extract their SFR from the Subfind Subhalo catalogs at different snapshots. For this, we only select subhaloes with a dark matter mass of $M_{\mathrm{dm}} > 0$ in order to guarantee that these galaxies are included in the merger trees catalogs. This gives $417$ and $4296$ subhaloes in the TNG50-1 and TNG100-1 run, respectively. Their SFT--mass relation is shown in Fig. \ref{fig:lcdmsft2}.

The third method cannot be applied for the EAGLE run because only 29 snapshots outputs (see table C.1 of \citet{McAlpine_2016} are publicly available and a fit cannot be made to get the SFT.

Given the test performed here which relies on quantifying the present-day (z = 0) galaxies in the cosmological simulations, the results unambiguously show that the models do not have the observed downsizing. This is in contradiction to other $\Lambda$CDM models (see section 2.3 of \citealt{2019MNRAS.487.3740W}) where they use their own quantification of measuring quenching times of model galaxies. All 3 methods result in a 5 sigma tension (Figs. \ref{fig:tngtau} and \ref{fig:lcdmsft2}).

\section{Collapse time}
\label{sec:collapsetime}
\begin{figure}
	\includegraphics[width=\columnwidth]{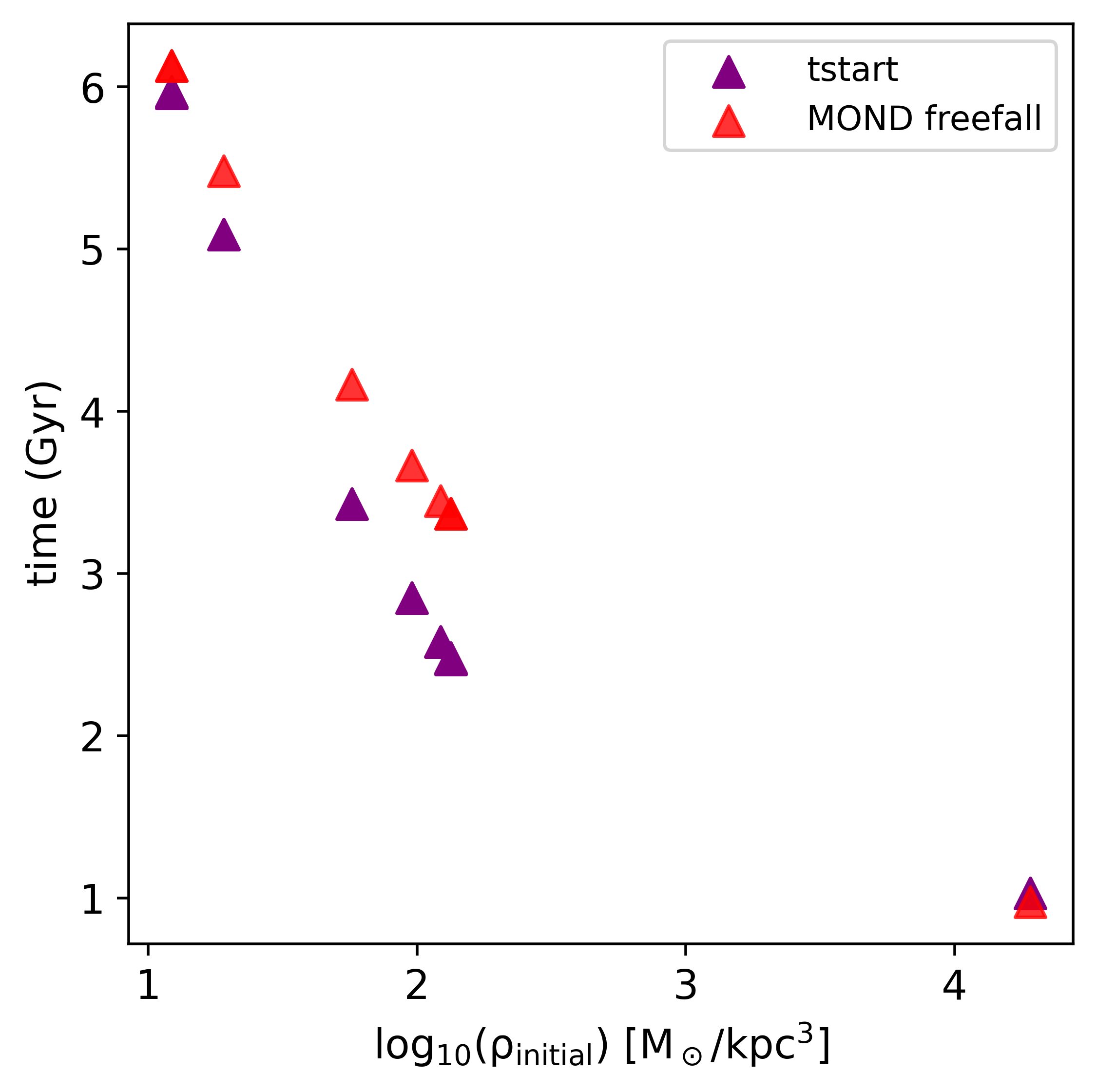}
    \caption{Collapse time -- cloud density relation for all models with $R_{\rm initial}$ = 500 kpc. The purple triangles are the time when the first stellar particle is formed in the simulation, $t_{\rm start}$ and the red triangles are the MOND free-fall time (Eq. \ref{eq:mondfreefall}).}
    \label{fig:mondfftime}
\end{figure}

The MOND free-fall time is the time taken for the gas cloud to collapse under its own gravitational attraction to form stars. This is calculated using Eq. \ref{eq:mondfreefall} (see \citealt{2021MNRAS.506.5468Z}). We compare the time taken for the first stellar particle to form in the simulation, $t_{\rm start}$, with this theoretical value (Fig. \ref{fig:mondfftime}). The $t_{\mathrm{start}}$, and the MOND free fall time are comparable in our simulation demonstrating consistency between the numerical and analytical results.

\section{Percentage change}
\label{sec:percentchange}
The percentage change (PC) quantifies the change from an initial value to the new value and expresses the change as an increase or decrease,

\begin{equation}
	\mathrm{PC} = \frac{(X_{2}-X_{1})}{|X_{1}|} \times 100.
	\label{eq:percentchange}
\end{equation}

Here, $X_{1}$ is the initial value and $X_{2}$ is the new value. If the PC obtained is a positive value then it is quantified as an increase from the initial value and if the obtained PC is negative then it is quantified as a decrease from the initial value. In Table \ref{tab:feedback}, the results obtained for the simple feedback (SF) models are used as the initial values and the results obtained for the complex feedback (CF) are used as the new values.
%\section{SFH of models in the high-density environment}
%The SFHs of the models in this work are represented in the high-density environment after shifting the peak of the SFH to the average-age defined by \citet{2005ApJ...621..673T} which shows that the first stellar particle forms earlier for more massive models (Fig. \ref{fig:sfhfourHD}).
%\begin{figure}
%	\includegraphics[width=\columnwidth]{figures/sfhsixHD.png}
%    \caption{Same as Fig. \ref{fig:sfhfour} but in high-density environment.}
%    \label{fig:sfhfourHD}
%\end{figure}

%%%%%%%%%%%%%%%%%%%%%%%%%%%%%%%%%%%%%%%%%%%%%%%%%%
% Don't change these lines
\bsp	% typesetting comment
\label{lastpage}
\end{document}